\documentclass[prb,onecolumn,nofootinbib,citeautoscript,10pt,notitlepage]{revtex4-2}
\usepackage{graphicx}			
\usepackage{amsmath}			
\usepackage{amsfonts}
\usepackage{amssymb}
\usepackage{amsthm}
\usepackage{xcolor}

\usepackage{color} 
\usepackage[papersize={8.5in,11in}]{geometry}
\usepackage[colorlinks=true]{hyperref}
\hypersetup{        
    unicode=false,          
    pdftoolbar=true,        
    pdfmenubar=true,        
    pdffitwindow=false,     
    pdfstartview={FitH},    
    pdfkeywords={keyword1} {key2} {key3}, 
    pdfnewwindow=true,      
    colorlinks=true,       
    linkcolor=magenta, 
    citecolor=blue,        
    filecolor=magenta,      
    urlcolor=blue           
} 

\usepackage[all]{hypcap}
\usepackage[caption=false]{subfig}

\newcommand{\beq}{\begin{equation}} 
\newcommand{\eeq}{\end{equation}}
\newcommand{\bqa}{\begin{eqnarray}} 
\newcommand{\eqa}{\end{eqnarray}}
\newcommand{\nn}{\nonumber \\}

\renewcommand{\vec}[1]{{\mathbf{#1}}}

\def\be{\begin{eqnarray}}
\def\ee{\end{eqnarray}}

\begin{document}

\title{Superconducting instability in non-Fermi liquids}

\author{Ipsita Mandal}
\affiliation{Perimeter Institute for Theoretical Physics, 31 Caroline St. N., Waterloo ON N2L 2Y5, Canada}

\begin{abstract}
We use renormalization group (RG) analysis and dimensional regularization techniques to study potential superconductivity-inducing four-fermion interactions in systems with critical Fermi surfaces of general dimensions ($m$) and co-dimensions ($d-m$), arising as a result of quasiparticle interaction with a gapless Ising-nematic order parameter. These are examples of non-Fermi liquid states in $d$ spatial dimensions. Our formalism allows us to treat the corresponding zero-temperature low-energy effective theory in a controlled approximation close to the upper critical dimension $d=d_c(m)$. The fixed points are identified from the RG flow equations, as functions of $d$ and $m$. We find that the flow towards the non-Fermi liquid fixed point is preempted by Cooper pair formation for both the physical cases of $(d=3, m=2)$ and $(d=2, m=1)$. In fact, there is a strong enhancement of superconductivity by the order parameter fluctuations at the quantum critical point.
\end{abstract}

\maketitle

\tableofcontents

\section{Introduction}
\label{intro}

Non-Fermi liquids are unconventional metallic states that cannot be studied using the framework of the Laudau Fermi liquid theory \cite{
holstein,reizer,leenag,HALPERIN,polchinski,ALTSHULER,YBKim,nayak,lawler1,lawler2,SSLee,metlsach1,metlsach,chubukov1,Chubukov,
chubukov3,mross,Jiang,Shouvik1,Lee-Dalid,shouvik2,ips1,ipshigh,ips4}. Such systems can arise when a gapless boson is coupled with a Fermi surface. One class of non-Fermi liquids involve the critical boson carrying zero momentum, such as the Ising-nematic critical point \cite{metlsach1,ogankivfr,metzner,delanna,delanna2,kee,lawler1,lawler2,rech,wolfle,maslov,quintanilla,yamase1,yamase2,halboth,
jakub,zacharias,YBKim,huh} and nonrelativistic fermions coupled with an emergent gauge field \cite{MOTRUNICH,LEE_U1,PALEE,MotrunichFisher}. In another class of non-Fermi liquids, the critical boson carries a non-zero momentum. The examples in this class include the order parameters at the spin density wave (SDW) and charge density wave (CDW) critical points \cite{metlsach,chubukov1,Chubukov,shouvik2}. All these non-Fermi liquids have a well-defined Fermi-surface but no well-defined Landau quasiparticles. In other words, although these metallic states do not have a discontinuity in the zero-temperature electron occupation number, the Fermi surface/momentum can be still be sharply defined through the location of weaker non-analyticities of the spectral function \cite{sudip,senthil}.

Superconducting instabilities in such non-Fermi liquid scenarios have attracted a lot of attention in the recent literature \cite{son,ips2,Max,ips3,sri}. We addressed this problem for nonrelativistic fermions coupled to a $U(1)$ gauge field in three spatial dimensions by solving the Dyson-Nambu equation \cite{ips2}. We used a similar approach to solve for the pairing gap for the cases of the half-filled Landau level state as well the bilayer Hall system with a total filling fraction equal to one \cite{ips3}. The cases of Ising-nematic critical point, spinon Fermi surface and the Halperin-Lee-Read states involving one-dimensional Fermi surfaces have been discussed in details by Metlitski \textit{et al}. \cite{Max} by using renormalization group (RG) techniques and the concept of ``inter-patch'' couplings. 

In the present work, we focus on the Ising-nematic quantum critical point involving a Fermi surface of generic dimensions, applying the RG methods developed in our earlier paper \cite{ips1}. Denoting the dimension of the Fermi surface by $m$  and the spatial  dimensions by $d$,
the co-dimension is given by $(d-m)$. It was shown that there is a non-trivial mixing between the ultraviolet (UV) and infrared (IR) physics for $m>1$, and a perturbative control for such systems can be achieved by tuning both $m$ and $(d-m)$ independently. This method also allows one to maintain the locality of the action in real space irrespective of the value of $m$, in contrast to the analysis by Metlitski \textit{et al}. \cite{Max}.

We consider a four-fermion interaction $V$ in the pairing channel with a tree-level scaling dimension equal to $-d+1+m/2$. We find that the scatterings in the BCS channel are enhanced by the square root of the Fermi surface volume, which goes as $k_F^m$. Consequently, the effective coupling that dictates the potential superconducting instability, is given by $ \tilde V = V k_F^{ m/2}$ , which has an enhanced scaling dimension of $ -d+1+ m$. Clearly, $\tilde V $ is marginal when the co-dimension takes the value $d-m =1 $. Our goal is to examine how the interactions of the fermions with critical bosons affect the pairing instability.

The paper is organized as follows:  In Sec.~\ref{review}, we review the dimensional regularization procedure devised in our earlier work to obtain a perturbative control of the Ising-nematic quantum critical point. The four-fermion interactions, which can be responsible for potential BCS instability, are introduced in Sec.~\ref{action}. There we compute the divergent terms contributing to the beta-functions to the lowest order. The solutions to the RG equations in various possible scenarios are discussed in Sec.~\ref{soln}. We conclude with a summary and outlook in Sec.~\ref{summary}.  The detailed computation of the relevant Feynman diagrams has been shown in the appendix.

\section{Ising-nematic quantum critical point}
\label{review}

\begin{figure}
\centering
\includegraphics[width=0.6 \textwidth]{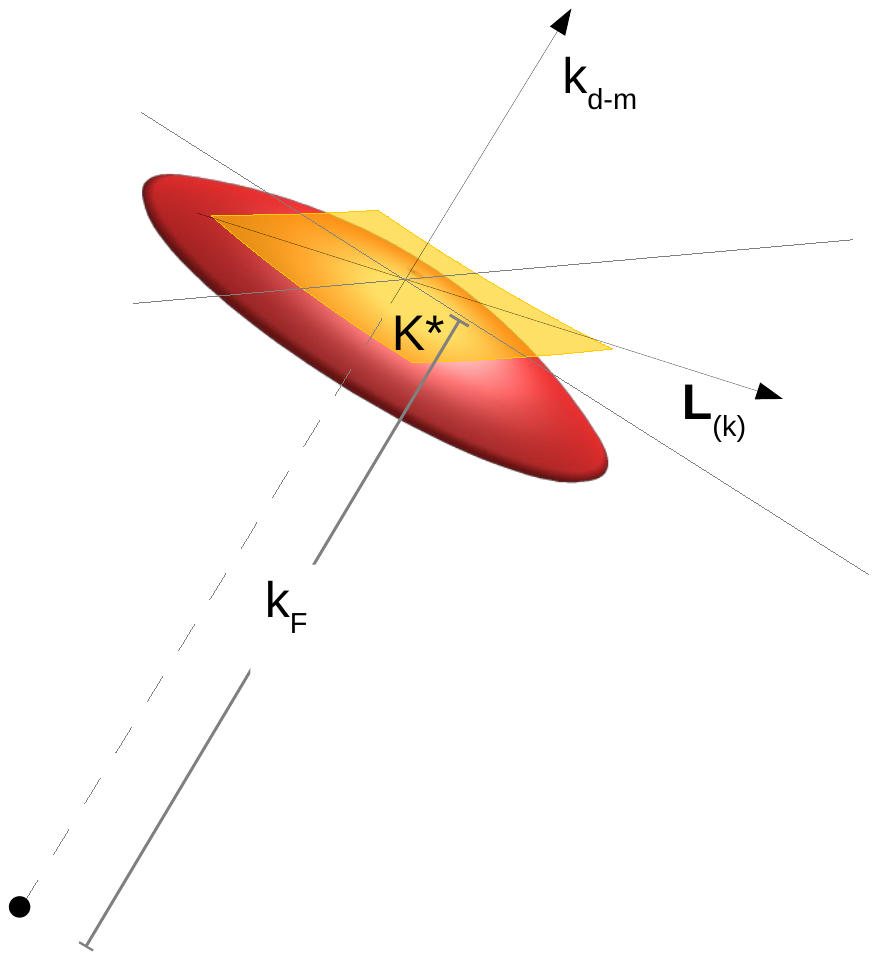} 
\caption{Cartesian coordinates on a patch of an $m$-dimensional convex Fermi surface.}
\label{fs-patch}
\end{figure} 

We consider the Ising-nematic quantum critical point involving an $m$-dimensional Fermi surface interacting with a massless boson in $d=(m+1)$ space dimensions \cite{ips1}, with coupling constant $e$. The momentum of the critical boson is centered at zero.
This system can be described by the action
\bqa
S & =&   \sum_{s=\pm,j}  \int dk \,
\psi_{s,j}^\dagger (k)
\Bigl[ 
 i \, k_0   +  s \, k_{1} +   {\vec L}_{(k)}^2  )  \Bigr] \psi_{s,j}(k) 
 + \frac{1}{2} \int  dk
 \left[ k_0^2 + k_{1}^2 +  {\vec L}_{(k)}^2 \right] \phi(-k) \, \phi(k) \nonumber \\
 && +  \frac{ e }{\sqrt{N}} \sum_{s=\pm,j}   
\int dk \, dq ~   \phi(q) ~  \psi^\dagger_{s,j}(k+q) \, \psi_{s,j}(k) \, ,
\label{act0-2}
\eqa
written in a coordinate system centred at an arbitrary point $K^*$ of the convex Fermi surface (see Fig.~\ref{fs-patch}).
Here, $k$ is the $(d+1)$-dimensional energy-momentum vector with
$dk \equiv \frac{d^{d+1} k}{(2\pi)^{d+1} }$. Assuming inversion symmetry,
we have included the ``right-moving" and ``left-moving" fermionic fields, $\psi_{+,j}$ and $\psi_{-,j}$, representing $K^*$ and its antipodal point $-K^*$.
These fields represent fermions
with flavours $j=1,2,..,N$, frequency $k_0$ and momentum $K_\alpha^*+k_\alpha $ ($-K_\alpha ^*+k_\alpha $) with $1 \leq \alpha \leq d$.
$k_{1}$ and  ${\vec L}_{(k)} ~\equiv~ (k_{2}, k_{3},\ldots, k_{d})$ 
denote the momentum components perpendicular and parallel to the Fermi surface at $\pm K^*$, respectively. 
The momenta are rescaled such 
that the Fermi velocity,
and the curvature of the Fermi surface at $\pm K^*$, 
are equal to one.

A perturbative control of the Yukawa coupling (and the strength of UV/IR mixing
in non-Fermi liquid states with $m>1$ \cite{ips1}) is achieved by
tuning the dimension \cite{Chakravarty,Fit,fit2,fit3,Torroba} and the co-dimension of the Fermi surface \cite{senshank,Lee-Dalid,shouvik2}, i.e.\ by writing
an action that describes an $m$-dimensional Fermi surface
embedded in the $d$-dimensional momentum space:
\begin{eqnarray}
S & = &  \sum_{j} \int dk \, \bar \Psi_j(k)
\Big [ 
i \, \vec \Gamma \cdot \vec K  
+ i \, \gamma_{d-m} \, \delta_k \Big ] \Psi_{j}(k) \, \exp \Big \lbrace \frac {{\vec{L}}_{(k)}^2}  { \mu \, {\tilde{k}}_F } \Big \rbrace 
+ \frac{1}{2} \int  dk \,
  {\vec{L}}_{(k)}^2\,  \phi(-k) \, \phi(k) \nonumber \\
 && +     \frac{i \, e \, \mu^{x/2}} {\sqrt{N}}  \sum_{j}  
\int  dk \,
\phi(q) \, \bar \Psi_{j}(k+q)\,  \gamma_{d-m} \Psi_{j}(k)\,,
\label{act5}
\end{eqnarray}
where a mass scale $\mu$ is introduced
to make ${\tilde {k}}_F$ and $e$ dimensionless. Here
\begin{equation}
x=-d +2 + \frac{m}{2} \,, \quad
\delta_k = k_{d-m} + \vec L _{(k)}^2\,, \quad 
k_F = \mu \, {\tilde {k}}_F \,, \quad \Psi_j^T(k) = \left( 
\psi_{+,j}(k),
\psi_{-,j}^\dagger(-k)
\right).
\end{equation}
The gamma matrices associated with $\vec K$ 
have been written as
$\vec \Gamma \equiv (\gamma_0, \gamma_1,\ldots, \gamma_{d-m-1})$. 
Since in real systems, $d-m$ lies between $1$ and $2$, 
we will consider only the $2 \times 2$ gamma matrices so that the corresponding spinors always have two components. We will use the representation
where
 \beq
 \Gamma_0  \equiv \gamma_0= \sigma_y \,, \quad \gamma_{d-m} = \sigma_x \,
 \eeq
are fixed in general dimensions.
The bare fermion propagator is given by 
$G_0 (k) =\frac{1}{i} \, \frac{\vec \Gamma \cdot \vec K +
\gamma_{d-m} \delta_k} 
{\vec K^2  + \delta_k^2} 
\exp \Big \lbrace - \frac {{\vec{L}}_{(k)}^2}  { k_F } \Big \rbrace$.
Note that we have used an exponential factor in order to damp out the fermion propagator, which acts as a soft cut-off for the directions tangential to the Fermi surface in loop-integrals. This restricts the integration along the directions tangent to the Fermi surface to a small region compared to $k_F$ by damping the propagation of fermions 
with $|{\vec{L}}_{(k)}| > k_F^{1/2}$. We denote the UV cut-off for $\vec K$ and $k_{d-m}$ as $\Lambda$, which is naturally given by the choice $\Lambda = \mu$.
Hence, two dimensionless parameters, $e$ and $\tilde k_F = k_F / \Lambda$, appear in the above action.
If $k$ is the typical energy 
at which we probe the system, we must impose the restriction $k << \Lambda << k_F$. 
The renormalization group flow is generated 
by changing $\Lambda$ and requiring that low-energy observables are independent of it.

The boson propagator that we will use in our computations includes the one-loop self-energy, and is given by
\bqa
\label{babos}
&& D_1(k) = \frac{1}{ {\vec{L}}_{(k)}^2 + \tilde \alpha \, 
\displaystyle\frac{   |\vec K|^{d-m}}{ |\vec{L}_{(k)}| } } \,,
\,  \, \tilde \alpha = \beta_{d} \, e^2 \mu^{x} k_F ^{ \frac{m-1}{2}} ,
\nn
&& \beta_d = \frac{  \Gamma^2 (\frac{d-m+1} {2})}
{2^{\frac{2d+m-1}{2}}  \pi^{\frac{d-1}{2}}  | \cos \lbrace  \frac{\pi (d-m+1)} {2} \rbrace |   \Gamma(\frac{d-m}{2}) \Gamma (d-m+1)}
\eqa
to the leading order in $k/k_F$,
for $|\vec K|^2/|\vec L_{(k)}|^2, ~\delta_k^2/|\vec L_{(k)}|^2 << k_F$. This is because the bare boson propagator is independent of $\left ( k_{0},\ldots ,k_{d-m} \right ) $, resulting in the loop integrations involving it being ill-defined. Hence we need to
resum a series of diagrams that provides a non-trivial dispersion along these directions \cite{ips1}.

In our previous work \citep{ips1}, it was established that
the higher order corrections are controlled not by $e$, 
but by an effective coupling ${\tilde e}$
with
\beq
{\tilde e} \equiv 
\frac{e^{ 2 (m+1 ) /3}  } {{\tilde{k}}_F ^{ \frac{(m-1) (2-m)} {6}}} \,.
\label{eq:eeff}
\eeq
The form of the one-loop fermion self-energy showed that it blows up logarithmically
in $\Lambda$ at the critical dimension
\beq
\label{dc}
d_c(m) = m + \frac{3}{m+1} \,.
\eeq
We also found that the RG equations,
with the increasing logarithmic length scale $l= - \ln \mu$, 
are given by
\bqa
 \frac{\partial \tilde k_F }{\partial l}  =  \tilde k_F  , \quad 
\frac{ \partial {\tilde e} }{ \partial l} = \frac{(m+1)\, \varepsilon}{3} \, {\tilde e} \,
- \frac{ (m+1)\,  u_1} {3N} \, {\tilde e}^2 \,,
\label{betae}
\eqa
to order ${\tilde e}^2$, where $\varepsilon = d_c -d $ and
\beq
\label{u1}
u_1 = \frac{ | \csc \lbrace (m+1) \pi/3 \rbrace |   }
{ 2^{m-1}   \pi^{\frac{m-2}{2}}
(4 \pi)^{\frac{3}{2(m+1)}}    \beta_{d}^{\frac{2-m}{3}}  (m+1) }
 \frac{  \Gamma (  \frac{m+4}{2(m+1)})  } 
{  \Gamma{(m/2)}  \Gamma(\frac{2-m}{2(m+1)})  \Gamma (  \frac{2m+5}{2(m+1)}) }\,.
\eeq


\section{Superconducting Instability}
\label{action}

With the motivation of analysing superconducting instability for our non-Fermi liquid system, we will add the appropriate four-fermion interaction terms to the action in Eq.~(\ref{act5}).
We note that:
\begin{eqnarray}
  \int \left ( \prod_{s=1}^4 {dp_s }  \right)
 &&\Big [
 \big \lbrace \bar{\Psi}_{j_1} (p_3) \, {\Psi}_{j_2} (p_1) \big \rbrace \, 
 \big \lbrace \bar{\Psi}_{j_3}  (p_4 ) \, {\Psi}_{j_4}  (p_2 ) \big \rbrace
-
 \big \lbrace \bar{\Psi}_{j_3}  (p_3)  \sigma_z  {\Psi}_{j_4} (p_1) \big \rbrace \, 
 \big \lbrace \bar{\Psi}_{j_1} (p_4 )  \sigma_z   {\Psi}_{j_2} (p_2 ) \big \rbrace
 \Big ] \nonumber\\
 && \times \,  \frac{    \mathcal{F}(p_1,p_3;p_2,p_4)  }   { 4 } \nn
=  -\int \left ( \prod_{s=1}^4 {dp_s }  \right)
 &&\Big [
\psi^{\dagger}_{+,j_1} (p_3)\,\psi^{\dagger}_{-,j_2} (- p_1)\, \psi_{-,j_3} (-p_4) \, \psi_{+,j_4} (p_2) 
\Big ] \,     \mathcal{F}(p_1,p_3;p_2,p_4)  
\,,\nn
\nonumber
\end{eqnarray}
can give rise to BCS pairing, where $ \mathcal{F}(p_1,p_3;p_2,p_4) $ is a function invariant under the simultaneous exchanges $(p_1, p_3) \leftrightarrow (p_2,p_4) $. Hence, we need to consider the four-fermion terms:
\bqa
\label{sc}
S^{\texttt{SC}}_{\texttt{gen}}
&=&  \frac{ \mu^{d_v}} { 4  }  \sum_{j_1,j_2,j_3,j_4 }  
\int \left ( \prod_{s=1}^4 {dp_s }  \right)
\,(2\pi)^{d+1} \, \delta^{(d+1)} (p_1+p_2-p_3-p_4 )  \nn
&& \qquad \qquad \times \,
\Big [
 \lbrace \bar \Psi_{j_1}(p_3) \Psi_{j_2 }(p_1) \rbrace  \, \lbrace \bar \Psi_{j_3 }(p_4)  \Psi_{j_4 }(p_2)\rbrace
-
 \lbrace \bar \Psi_{j_3 }(p_3) \sigma_z  \Psi_{j_4 }(p_1) \rbrace  \, 
 \lbrace \bar \Psi_{j_1 }(p_4)  \sigma_z  \Psi_{j_2 }(p_2)\rbrace
 \Big ] \,,\nn
 && 
 \qquad \qquad \times \, 
 \Big[
  V_{S}(\vec {p}_1,\vec{p}_3; \vec {p}_2,\vec {p}_4) \left( \delta_{j_1,j_3 } \delta_{j_2,j_4 } -  
  \delta_{j_1,j_4} \delta_{j_2,j_3 }\right)
  +
  V_{A}(\vec {p}_1,\vec{p}_3; \vec {p}_2,\vec {p}_4) \left( \delta_{j_1,j_3 } \delta_{j_2,j_4 } +  
  \delta_{j_1,j_4} \delta_{j_2,j_3 }\right)
 \Big] \nn
 &&
\eqa
where $\vec k $ denotes the momentum components, $(k_{d-m}, \vec{L}_{(k)})$, of the $(d+1)$-dimensional energy-momentum vector $k$. The subscript ``$S$" (``$A$") denotes that $V_S(\vec {p}_1,\vec{p}_3; \vec {p}_2,\vec {p}_4)$ $\left ( V_A(\vec {p}_1,\vec{p}_3; \vec {p}_2,\vec {p}_4) \right ) $ is symmetric (antisymmetric) under the interchange $ 
p_1 \leftrightarrow p_3$ or $ p_2 \leftrightarrow p_4 $. Here we have have made the $V_{S/A}$'s dimensionless by using the mass scale $\mu$, with $d_v = -d+1+ \frac{m}{2}$ representing the scaling dimension of $V_{S/A}$.

\begin{figure}[t]
\centering
\subfloat[][]
{\includegraphics[width=0.32 \textwidth]{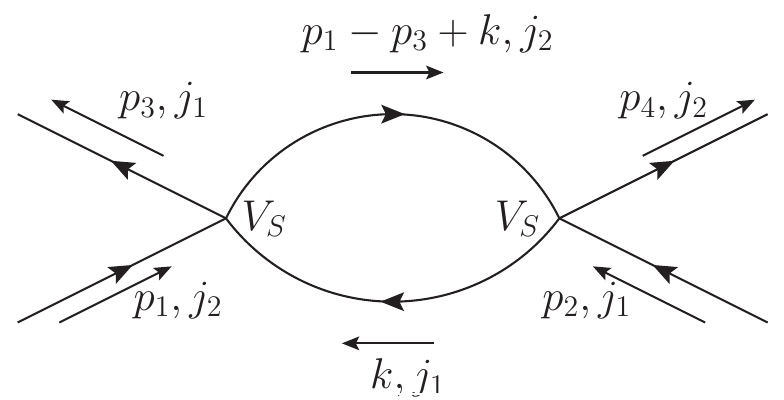} 
{ \label{fig:4vv1}}}
\subfloat[][]
{\includegraphics[width=0.35 \textwidth]{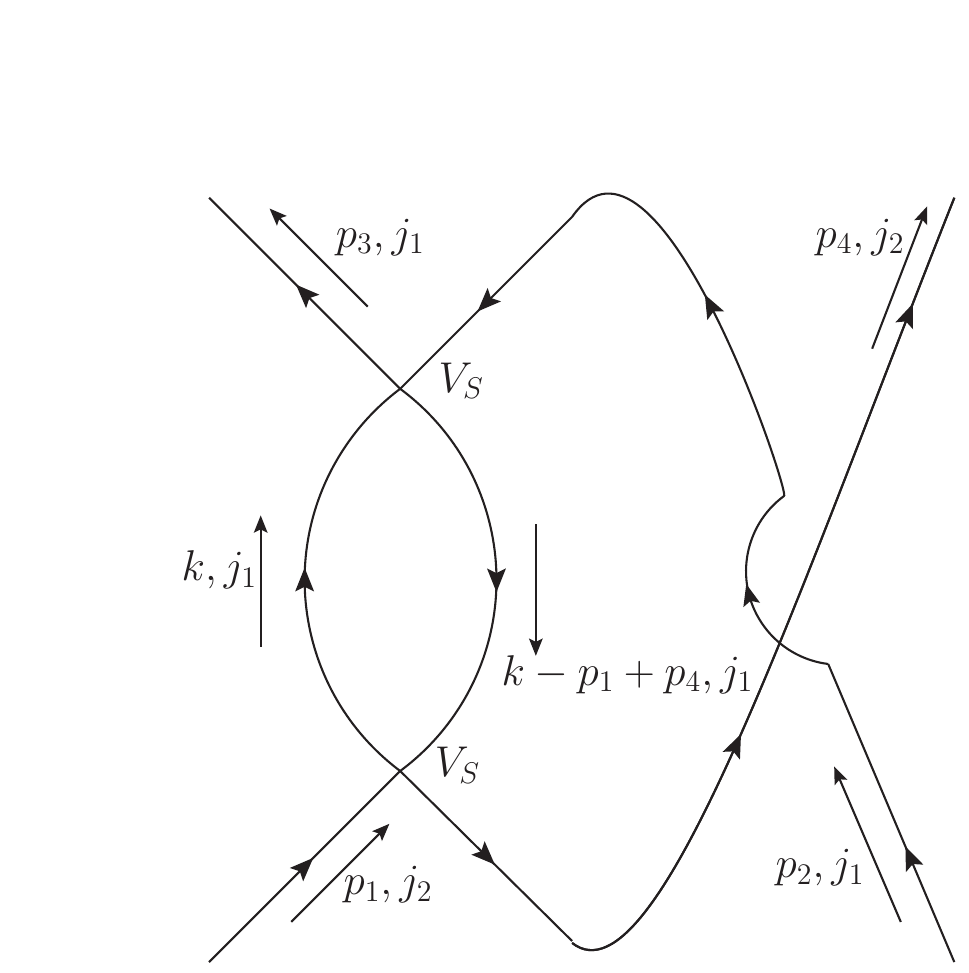}
{\label{fig:4vv2}}}
\subfloat[][]
{\includegraphics[width= 0.26 \textwidth]{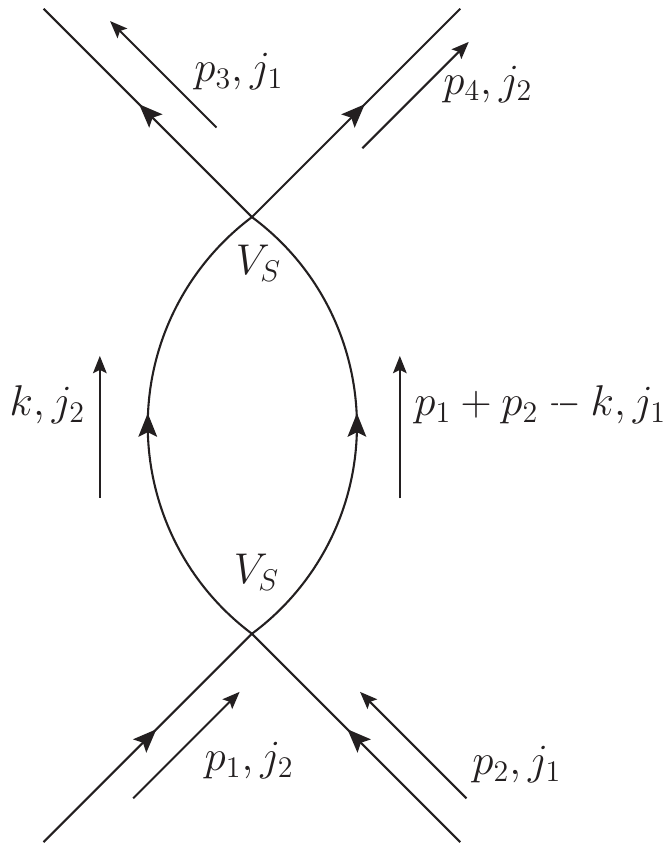}
{\label{fig:4vv3}}}
\caption{\label{fig:loop-4VV}One-loop diagrams proportional to $V_{S}^2.$ Here $p_4=p_1+p_2-p_3$.}
\end{figure}

We will find from our analysis of the RG flow, in order for superconducting instability to set in, we must have $\vec p_1=\vec p_3$ and $\vec p_2 = \vec p_4 $.
Furthermore, assumption of rotational symmetry allows us to write:
\beq
V_{S/A}(\vec {p}_1,\vec{p}_1; \vec {p}_2,\vec {p}_2) =
V_{S/A}(\theta_1- \theta_2) ,
\eeq
where $\theta_{1,2}$ denote the angles for $\vec{p}_{1,2} $ on the Fermi surface. For simplicity and in order to obtain analytic expressions, we will consider only the case of a constant non-zero $V_S $ , for which we need at least two flavours of fermions. Then, with $N=2$, Eq.~(\ref{sc}) reduces to:
\bqa
\label{scaction}
S^{\texttt{SC}}
&=& -\frac{ \mu^{d_v}  V_{S}} { 4  }  \sum_{j_1,j_2 }  
\int \left ( \prod_{s=1}^4 {dp_s }  \right)
\,(2\pi)^{d+ 1} \, \delta^{(d+1)} (p_1+p_2-p_3-p_4 )  \,
\left( 1-  \delta_{j_1,j_2 } \right)\nn
&& \qquad \qquad \times \,
\Big [
 \lbrace \bar \Psi_{j_1}(p_3)   \Psi_{j_2 }(p_1) \rbrace  \, \lbrace \bar \Psi_{j_2  }(p_4)   \Psi_{j_1 }(p_2)\rbrace
 -
 \lbrace \bar \Psi_{j_1 }(p_3) \sigma_z  \Psi_{j_2 }(p_1) \rbrace  \, 
 \lbrace \bar \Psi_{j_2 }(p_4)  \sigma_z  \Psi_{j_2 }(p_1)\rbrace
 \Big ] \,,\nn
\eqa
It is straightforward to apply our formalism for a system with a given number of flavours and a specific angular dependence
of $V_{S/A}$.

 \subsection{One-loop diagrams generating terms proportional to $V_{S}^2$ }

 We need to consider one-loop diagrams as shown in Fig.~\ref{fig:loop-4VV} for contributions proportional to  $V_{S}^2$. The computation of these diagrams has been detailed in
 Appendix~\ref{v2}.
The contribution from Fig.~\ref{fig:4vv1} is found to be proportional to:
\begin{eqnarray}
- (1 -\delta_{j_1,j_2})^2 \int dk \, \mbox{Tr} \big[ G_0 (k+ \mathbf{P_1} - \mathbf{P_3}) \, G_0 (k)\big]
= \frac{  2^{ 1 -2d +\frac{m}{2} } 
\left ( 1- \delta_{j_1,j_2}  \right ) \, k_F^{m/2}  \, \pi^{ 1-\frac{d} {2} }   \, \sec \left ( \frac{ (d-m)\, \pi } {2} \right )
} 
{ \Gamma \left ( \frac{ d-m} { 2 } \right )   
\, | \mathbf{  P_3 -P_1} |^{-d+m+ 1}} \,,\nn
&&
\end{eqnarray}
which is logarithmically divergent at $d-m=1$ such that we can express it as $\frac{ k_F^{m/2} \,\ln \left( \frac{\Lambda} {|\mathbf{P_1} - \mathbf{P_3} |}\right) }
{2^ { \frac{3m}{2}} \, \pi^{ 1 + \frac{m}{2}} }$.
The results from Figs.~\ref{fig:4vv2} and \ref{fig:4vv3} are $k_F$-suppressed and hence do not contribute to the beta-functions.

Noting that $\mbox{Tr} \big[ \sigma_z\, G_0 (k+ \mathbf{P_1} - \mathbf{P_3})\,\sigma_z \, G_0 (k)\big]
= - \mbox{Tr} \big[ G_0 (k+ \mathbf{P_1} - \mathbf{P_3}) \, G_0 (k)\big] $, the full contribution from all one-loop diagrams proportional to $V_{S }^2$ is given by:
\begin{align}
& t_{VV} =
 \frac{4\times 2}{2 !} \frac{4}{ 4 \times 4  } \times \frac{ 2^{ 2 + 2d-\frac{m}{2}} \, k_F^{m/2}  \, \mu^{2 \, d_v}
 \, V_{S}^2 }
{  \pi^{ \frac{d}{2}}   \,\Gamma \left ( \frac{ d-m} { 2 } \right )  \epsilon} + \mathcal{O}\left( \epsilon\right)  
= \frac{ v_2 \, \mu^{2 \, d_v +\frac{m} {2} } \,\tilde{k}_F^{m/2} \, V_{S}^2 }{\epsilon}  + \mathcal{O}\left( \epsilon\right)  \,,\nn
& v_2 =  \frac{ 2^{ 2 -2d+\frac{m}{2}}  }
{ \pi  ^{ \frac{d}{2}  }  \,  \Gamma \left ( \frac{ d-m} { 2 } \right ) } \,,
\end{align}
for $d-m=1-\epsilon$.

\subsection{UV-divergent terms proportional to $ \tilde e $ }

Following earlier studies \cite{son,Max}, we have to include contributions from tree-level BCS channel, which are generated from long-range interactions between the fermions. In this case, the long-range interaction is mediated by the massless boson. Here we treat this contribution in the language of dimensional regularization as we have done for the other terms. In Appendix~\ref{wil}, we have provided an alternative treatment of these terms in the more familiar Wilsonian RG language.

\begin{figure}
        \centering
                \includegraphics[width=0.35 \textwidth]{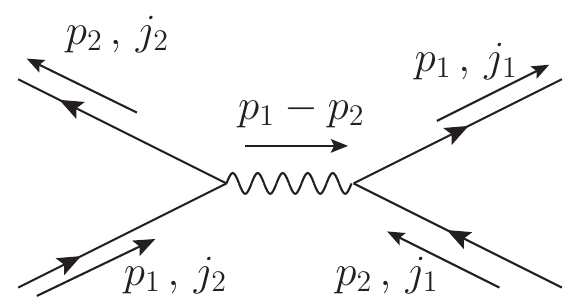} 
                \caption{Tree-level diagram proportional to $e^2.$ }
\label{bcstree}
\end{figure} 

We consider Fig.~\ref{bcstree}, which gives the terms 
\begin{eqnarray} 
&& \sum_{j_1, j_2} \Big [ 
- \psi^{\dagger}_{+,j_1 } (p_2)\,\psi^{\dagger}_{-,j_2} (- p_2)  \, \psi_{-,j_2} (-p_1) \, \psi_{+,j_1} (p_1)
-  \psi^{\dagger}_{+,j_1 } ( p_1 ) \, \psi^{\dagger}_{-,j_2 } ( -p_1)
     \, \psi_{-,j_2 } (-p_2)  \, \psi_{+,j_1 } (p_2)   \nn
&& 
 \quad \quad  - \, \psi^{\dagger}_{+,j_1} (p_2) \, \psi_{+,j_1 } (p_1 ) \,  \psi^{\dagger}_{+,j_2 } (p_1) \, \psi_{+,j_2 } (p_2 ) 
 -   \psi^{\dagger}_{-,j_2} ( -p_2)\, \psi_{-,j_2} ( -p_1 )
\,\psi^{\dagger}_{-,j_1 } (-p_1)  \, \psi_{-,j_1 } (- p_2 )
  \Big ]  \nonumber
\end{eqnarray}
multiplying $ \frac{(ie)^2  \mu^x   D_1(p_1-p_2) } {2 \, N}  $. Due to symmetry of $D_1(p_1-p_2)$ under $ p_1 \leftrightarrow p_2 $, they contribute as $     \frac{ e^2  \mu^x   D_1(p_1-p_2) } { N}  $ for terms responsible for BCS instability.

Using $   \mathbf{L}_{(q)} ^ 2   \simeq 2\, k_F^2 (1- \cos \theta)/k_F\Rightarrow 
 |\mathbf{L}_{(q)}| \simeq  \sqrt{k_F} \, |\theta|$, the decomposition into angular momentum channels for an $m$-dimensional Fermi surface leads to the contribution being proportional to:
 \begin{eqnarray}
t_{ee}
&\simeq &
  \frac{ e^2 \, \Lambda^x} {2 \, N} \times 2  \int_{\theta >0} 
   { d \theta}   \, \frac{ \theta^{m-1} \, |\mathbf{L}_{(q)}|} 
   { \mathbf{L}_{(q)} ^ 3  + \tilde \alpha \, |\vec Q |^{d-m} } 
=    \frac{  e^2 \, \Lambda^x} {  N\, k_F^{m/2}  }  \int_{\theta >0} 
   { d  |\mathbf{L}_{(q)}|}   \, \frac{ |\mathbf{L}_{(q)}|^ m } 
   { \mathbf{L}_{(q)} ^ 3  + \tilde \alpha \, |\vec Q |^{d-m} } \nn
&=&
  \frac{ e^2 \, \Lambda^x  \,   \Gamma \left( \frac{m+1} {3}\right ) \, \Gamma \left( \frac{ 2-m} {3}\right )} 
 {3 \, N\, k_F^{m/2} \, \tilde \alpha^{\frac{2-m}{3} }  \, |\vec Q |^{ \frac{(d-m)(2-m)}{3} }
  } \,,
\end{eqnarray}
which is independent of the angular momentum channel.
This expression tells us that for $m= 2 - \delta $ and $d= m+1-\gamma \, \varepsilon $, we get a pole such that
$t_{ee} \simeq   \frac{  \tilde e \, \Lambda^{ \gamma \, \varepsilon+
\frac{ ( 2-m )\, (m-1) } {6 }  }  
\,   \Gamma \left( \frac{m+1} {3}\right ) } 
 { \beta_d^{\frac{ 2-m }{3}}   N\, k_F^{m/2}} \,
 \frac { 1} {\delta } $,
which is equivalent to the logarithmically divergent term $   \frac{ e^2 \, \Lambda^x  \,   \Gamma \left( \frac{m+1} {3}\right ) } 
 {  N\, k_F^{m/2}} 
 \ln \left (  \frac{\sqrt{ k_F}}  {\tilde \alpha^{\frac{1}{3}} \, |\mathbf{Q}|^{\frac{d-m}{3}} }  \right ) $. This term, when continued to the strongly coupled regime of $m=1$, will translate into the infrared divergence there. The resulting counterterm is given by 
\bqa
\label{v1}
&& - \frac{    \mu^{\lambda_1   \varepsilon} \,  \tilde{ e } \, v_1 } { 4 \, N\, a \, \epsilon }  \sum_{j_1,j_2 }  
\int \left ( \prod_{s=1}^4 {dp_s }  \right)
 (2\pi)^{d+1 } \, \delta^{(d+1)} (p_1+p_2-p_3-p_4 )  
\left ( 1-  \delta_{j_1, j_2 }   \right)  \nn
&& \qquad \qquad \qquad  \times \,
\Big [
 \lbrace \bar \Psi_{j_1}(p_3)   \Psi_{j_2 }(p_1) \rbrace  \, \lbrace \bar \Psi_{j_1 }(p_4)    \Psi_{j_2 }(p_2)\rbrace
-
 \lbrace \bar \Psi_{j_1 }(p_3)  \sigma_z  \Psi_{j_2 }(p_1) \rbrace  \, 
 \lbrace \bar \Psi_{j_1 }(p_4)   \sigma_z  \Psi_{j_2 }(p_2)\rbrace
 \Big ]  ,\nn
&&  \lambda_1 =1-  \frac{a\,(7-m^2)} { 6 \, (m+1) } \,,\quad
a \, \epsilon = 2-m\, , \quad
v_1 = \frac{    \Gamma \left( \frac{m+1} {3}\right )}
{ \beta_d^{\frac{ 2-m }{3} }}  \,.
\eqa

 \subsection{One-loop diagrams generating terms proportional to $\tilde e \, V_{S} $ }

\begin{figure}
        \centering
        \subfloat[][]
                {\includegraphics[width= 0.3  \textwidth]{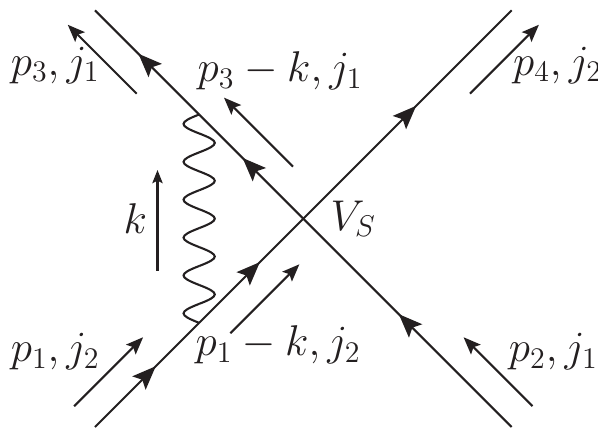}
               { \label{fig:4V1}} }
        \subfloat[][]
                {\includegraphics[width= 0.3  \textwidth]{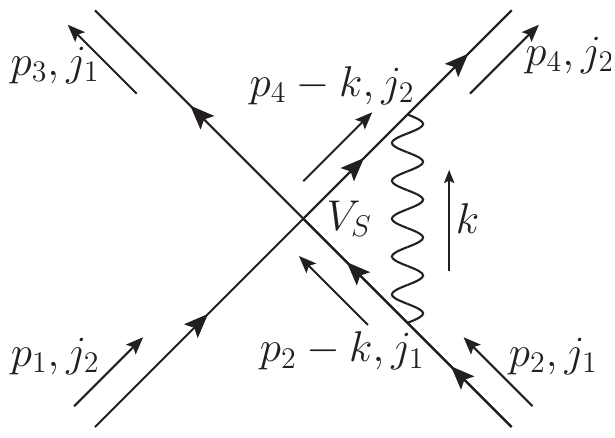}
		{\label{fig:4V2}}} 
\subfloat[][]{\includegraphics[width= 0.3  \textwidth]{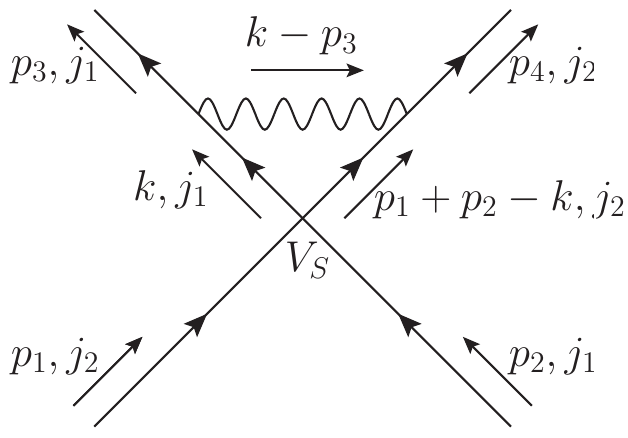}
                {\label{fig:4V3}} }
                
        \subfloat[][]
               { \includegraphics[width= 0.3  \textwidth]{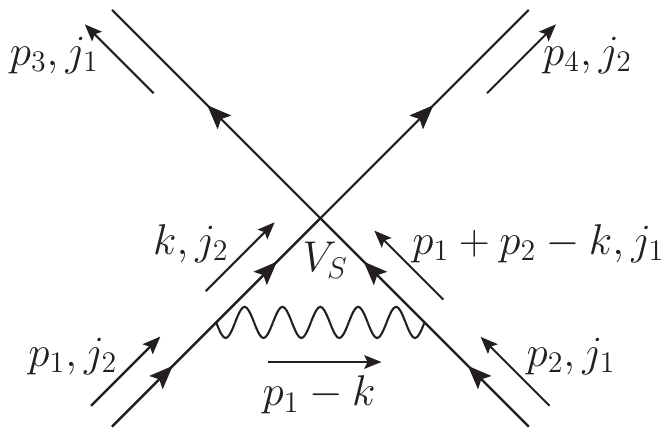}
                {\label{fig:4V4}} }
        \subfloat[][]
                {\includegraphics[width= 0.3  \textwidth]{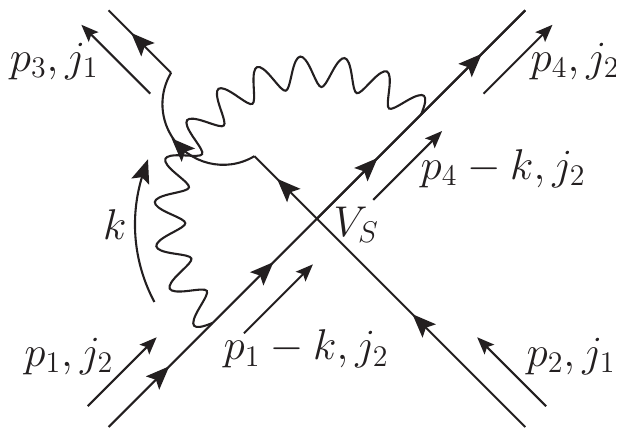}
                 {\label{fig:4V5}} }
        \subfloat[][]
                {\includegraphics[width=0.3 \textwidth]{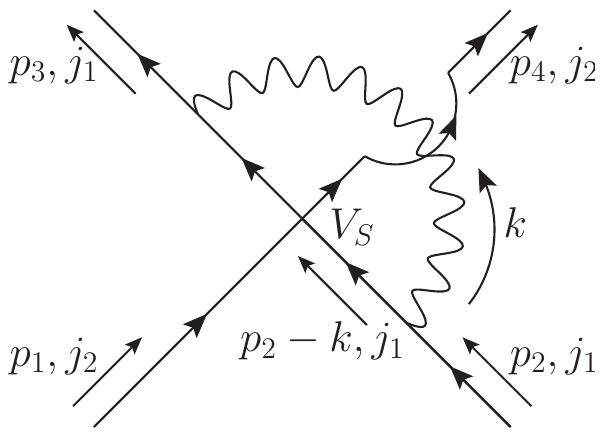} {\label{fig:4V6}}    }  
\caption{\label{fig:loop-4V}One-loop diagrams proportional to $ \tilde e \, V_{S}$, each consisting of two fermionic and one bosonic propagators forming the loop. Here $p_4=p_1+p_2-p_3$, $\vec p_1 = \vec p_3$ and $\vec p_2 = \vec p_4$.
.}
\end{figure}

\begin{figure}
        \centering
       \subfloat[][]
               { \includegraphics[width= 0.37   \textwidth]{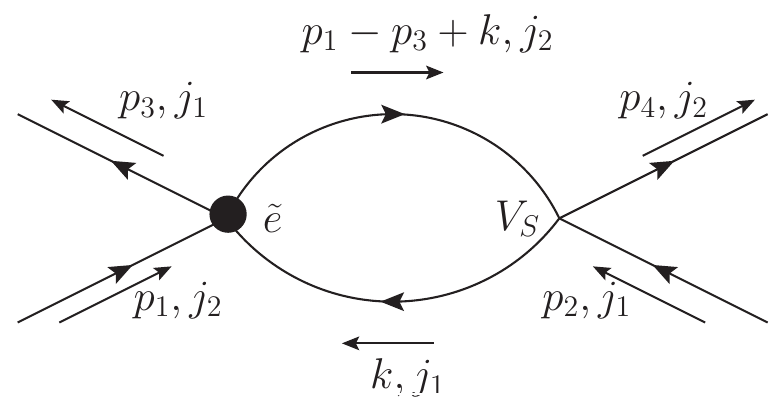}
             { \label{eV1}} }
       \subfloat[][]
               { \includegraphics[width= 0.37 \textwidth]{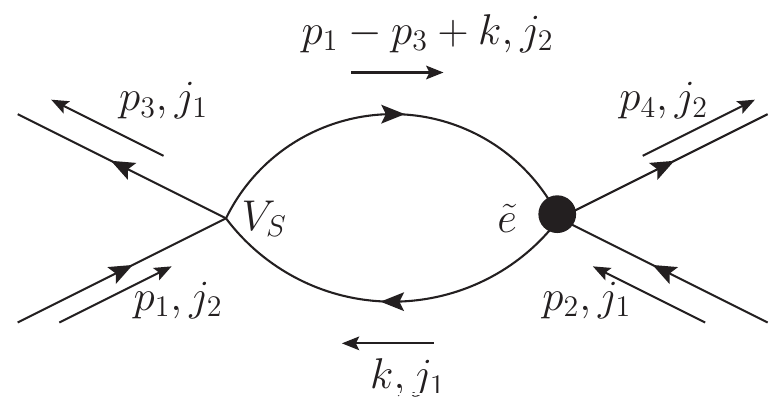}
		{\label{eV2}} }
%
\caption{\label{evc}One-loop diagrams proportional to $ \tilde e \, V_{S}$, resulting from counterterms. The counterterm vertex has been denoted by a blob. Here $p_4=p_1+p_2-p_3$.}
\end{figure}

 The set of one-loop diagrams which can generate terms proportional to $\tilde e \, V_{S} $  are shown in Fig.~\ref{fig:loop-4V}. The computation of these diagrams has been detailed in
 Appendix~\ref{v1l}. There it has been found that only Figs.~\ref{fig:4V1} and  \ref{fig:4V2} contribute. After cancellation with the appropriate diagrams in Fig.~\ref{evc} consisting of counterterm vertices, as explained in Appendix~\ref{logsq}, their contribution to the beta-function is captured by the term
\bqa
\left (  \frac{v_3  }   {a}  +  v_4  \right)
\frac{ \tilde e \,V_S \,  \mu^{\left (\lambda_2 + \gamma \right )\, \varepsilon}   }
{  N \, \varepsilon } \,,
 \eqa
 where
 \begin{eqnarray}
 v_3 =\frac{2\, \gamma_E }
 { \beta_d^{  \frac{2-m} {3}  } \,(2   \pi )^{\frac{3} {5-m}+m-2  }
 \, \Gamma \left (  \frac{m} {2}  \right)   \Gamma \left (  \frac{3} {2  \left(  m+1 \right ) }  \right) \pi
 } \,,
 \quad
 v_4=\frac{3 \,  v_3 } { m+1 } - \frac{ 2 } {3} \,, \quad
  \lambda_2 =  \frac{ a\,(m-1) }    {6}  \,.\nn
 \end{eqnarray}
 Furthermore, the contribution from Figs.~\ref{eV1} and \ref{eV2}, after cancellation with Figs.~\ref{fig:4V1} and \ref{fig:4V2}, is given by: 
\bqa
\left (  \frac{v_5  }   {\gamma }  + \frac{ v_6 } {a}  \right)
\frac{ \tilde e \,V_S \,  \mu^{ \lambda_1 \varepsilon}   }
{  N \, \varepsilon } \,,
 \eqa
 where
 \begin{eqnarray}
 v_5 =- \frac{2^{5 -2d+\frac{m} {2} } \, \gamma_E \,  \Gamma \left (  \frac{m+1} {3}  \right)
  }
 { \beta_d^{  \frac{2-m} {3}  } \,    \pi ^{\frac{d} {2}   }
 \, \Gamma \left (  \frac{d-m} {2}  \right)   
 } \,,
 \quad
 v_6=- v_5   \,.\nn
 \end{eqnarray}

\begin{figure}
\centering
 \subfloat[][]
               { \includegraphics[width= 0.4   \textwidth]{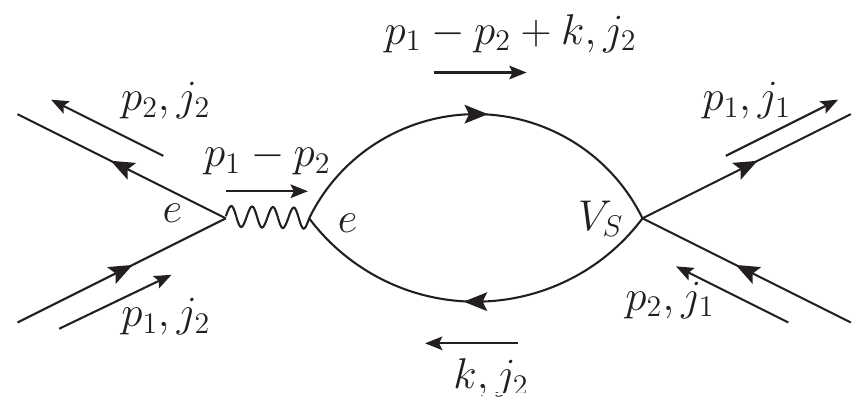}
             { \label{eVzero1}} }
\subfloat[][]
               { \includegraphics[width= 0.4  \textwidth]{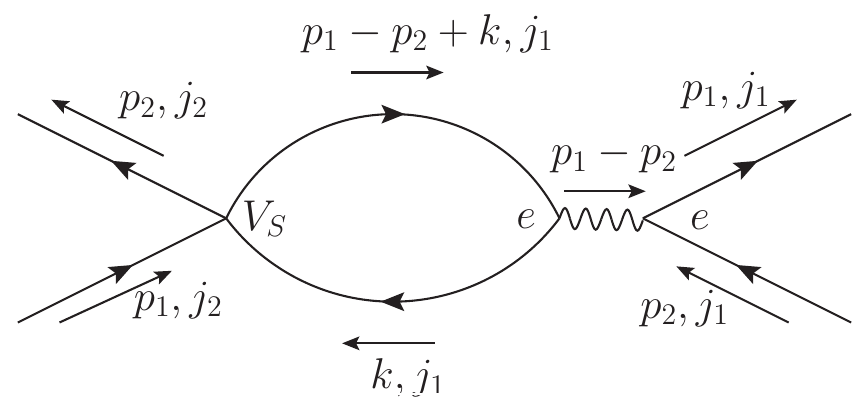}
		{\label{eVzero2}} }  
\caption{\label{evzero}One-loop diagrams proportional to $ e^2 \, V_{S}$, each consisting of two Yukawa vertices and one fermionic loop.}
\end{figure}

There are two more diagrams in this class, as shown in Fig.~\ref{evzero}. Due to the vanishing of the trace of the gamma matrices in the fermionic loop, they are identically zero.

 \subsection{ Beta-functions for the coupling constants}
 
\label{betafn}

We have found that the one-loop divergent terms are proportional to $\frac{ {k}_F^{\frac{m}{2} }  \mu^{2 d_v  } }  {\gamma \, \varepsilon}  \, V_S^2$ for $d-m=1 -\gamma \, \varepsilon \,$, for  $ |\gamma \, \varepsilon| \ll 1$. Hence, the upper critical dimension for the four-fermion BCS interaction is $\tilde d_c=m+1$, which is different from $d_c = m+\frac{3}{m+1}$ for the Yukawa coupling (see Eq.~(\ref{dc})). For $m=2$, $d_c = \tilde d_c = 3$. However, for $m=1$, $d_c = 5/2$ and $\tilde d_c = 2 $.
In order to study both the physically relevant cases of $m=1$ and $m=2$, we define $\gamma$ and $\varepsilon$ such that
\begin{equation}
d_c -\varepsilon = \tilde d_c -\gamma \, \varepsilon
\quad  \Rightarrow \,
\gamma \, \varepsilon= \varepsilon - \frac{ 2-m } {m+1}   \,.
\end{equation}
Furthermore, from the form of the divergent terms, it is apparent that
\beq
\tilde{V}_{S} = \tilde{k}_F^{m/2} \, V_{S} 
\eeq
is the effective coupling constant for the four-fermion interactions, which has an enhanced tree-level scaling dimension given by $d_v+ 1/2 =\tilde d_c -d $. $\tilde{V}_{S}$ is marginal when the co-dimension $d-m=1 $.

The counterterms for the four-fermion interaction term take the form:
\bqa
\label{actsc-CT}
S^{\texttt{SC}}_{ct} &=&  - \frac{    \mu^{\gamma \, \varepsilon} \tilde{ A }_S \tilde{ V}_S} { 4  }  \sum_{j_1,j_2 }  
\int \left ( \prod_{s=1}^4 {dp_s }  \right)
 (2\pi)^{d+1 } \, \delta^{(d+1)} (p_1+p_2-p_3-p_4 )  
\left (  1- \delta_{j_1, j_2 }    \right)  \nn
&& \qquad \qquad \times \,
\Big [
 \lbrace \bar \Psi_{j_1}(p_3)   \Psi_{j_2 }(p_1) \rbrace  \, \lbrace \bar \Psi_{j_1 }(p_4)    \Psi_{j_2 }(p_2)\rbrace
-
 \lbrace \bar \Psi_{j_1 }(p_3)  \sigma_z  \Psi_{j_2 }(p_1) \rbrace  \, 
 \lbrace \bar \Psi_{j_1 }(p_4)   \sigma_z  \Psi_{j_2 }(p_2)\rbrace
 \Big ]  ,\nn
\eqa
where we use the definition
\bqa
\tilde{A}_{S} 
\equiv  \tilde{Z}_S -1
=  \sum_{ \alpha_1   =1}^\infty \frac{\tilde{Z}_{S,\alpha_1 }(\tilde e,\tilde{k}_F, V_S )}
{    \varepsilon ^{\alpha_1}   } \,.
\eqa

Adding the counterterms to the original $S^{\texttt{SC}} $, and denoting the bare quantities by the superscript ``$B$",
we obtain the renormalized four-fermion interaction as:
\bqa
\label{actsc-ren}
S^{\texttt{SC}}_{\texttt{ren} } &=&     -  \frac{  V_S^{\texttt{B} } }  { 4 }  \sum_{j_1,j_2 }  
\int \left ( \prod_{s=1}^4 {dp^{\texttt{B} }_s }  \right)
\,(2\pi)^{d+1 } \, \delta^{(d+1)} (p^{\texttt{B} }_1+p^{\texttt{B} }_2-p^{\texttt{B} }_3-p^{\texttt{B} }_4 ) 
\left ( 1-  \delta_{j_1, j_2 }  \right) \, \nn
&& \qquad \qquad \times \,
\Big [
 \lbrace \bar \Psi^{\texttt{B} }_{j_1}(p_3) \Psi^{\texttt{B} }_{j_2 }(p_1) \rbrace  \, \lbrace \bar \Psi^{\texttt{B} }_{j_1 }(p_4)     \Psi^{\texttt{B} }_{j_2 }(p_2)\rbrace
 -
 \lbrace \bar \Psi^{\texttt{B} }_{j_1 }(p_3)  \sigma_z  \Psi^{\texttt{B} }_{j_2 }(p_1) \rbrace  \, 
 \lbrace \bar \Psi^{\texttt{B} }_{j_1 }(p_4)  \sigma_z  \Psi^{\texttt{B} }_{j_2 }(p_2)\rbrace
 \Big ] \,,\nn
\eqa
where
\bqa
\vec K & = & \frac{Z_2}{Z_1} \vec K_B \, , \quad k_{d-m} = k_{B, d-m} \, , \quad {\vec{L}}_{(k)} = {\vec{L}}_{(k), B} \,, \quad
\Psi(k) =  Z_\Psi^{-1/2} \,\Psi_B(k_B)\,, \quad \phi(k) = Z_\phi^{-1/2} \phi_B(k_B)\,, \nn
e_B & = & Z_3^{-1/2} \left( \frac{Z_2}{Z_1} \right)^{(d-m)/2} \mu^{ x/2} \, e \,,\quad   k_F =\mu \, {\tilde{k}}_F \,,
\quad
 V_S^{\texttt{B} } \, k_F^{m/2} = \tilde{Z}_S \, Z_{\psi}^{-2}  \left( \frac{Z_2}{Z_1} \right)^{3 (d-m)} \mu^{  \gamma \, \varepsilon} \, \tilde{V}_S \,.
\eqa

The dynamical critical exponent and the anomalous dimension for the fermions are given by \cite{ips1}:
\bqa
z= 1+ \frac{m+1} {3 \, N} \, u_1 \, \tilde e , \quad
\eta_\psi = (1-z) (d-m)/2 \,,
 \eqa
with $u_1$ defined in Eq.~(\ref{u1}).
%
%
Using the definition $2-m = a \, \varepsilon \Rightarrow a= \frac{ 3 \,(1 - \gamma )  } { 1 \, + \, (1- \gamma )
\, \varepsilon } $, we have
\begin{eqnarray}
&& \frac{ \tilde{Z}_{S, 1 }  \, \tilde{V}_S \, \mu^{\gamma \, \varepsilon } }
{    \varepsilon  }
=  \frac{ \left (  \frac{ v_1 } {a}   + \frac{v_5 \tilde{V}_S} {\gamma }  + \frac{ v_6 \tilde{V}_S } {a} \right ) 
\tilde{e} \, \mu^{\lambda_1 \, \varepsilon } } {N   \, \varepsilon }
+ \frac{v_2 \, \tilde{V}_S^2 \, \mu^{\gamma \, \varepsilon} } { \gamma \, \varepsilon }
+\frac{ \left(   \frac{v_3 } {a } +    v_4  \right) 
              \tilde e \, \tilde V_S \, \mu^{\left (\lambda_2 +\gamma \right )  \varepsilon }} 
{ N \, \varepsilon}\,.
\end{eqnarray}
Implementing the condition $\frac{d}{d \ln \mu}\left(  V_S^{\texttt{B} } \, k_F^{m/2} \right) =0 $, we get
\begin{align}
&
\beta_V 
+\frac{ \Big \lbrace   \frac{ v_1 } {a}  +  \left ( \frac{v_3 +v_6 } {a } +    v_4  
+ \frac{v_5  } {\gamma }    \right ) \tilde{V}_S
\Big \rbrace  \, \tilde{\beta_e} +
\left ( \frac{v_3 + v_6 } {a } +    v_4 +  \frac{v_5  } {\gamma }   \right ) \, \beta_V \, \tilde e  } 
{ N \, \varepsilon} 
 + \frac{   2 \,v_2 \, \tilde V_S \, \beta_V
 } 
 {\gamma \, \varepsilon}  
+ \left \lbrace  \gamma \, \varepsilon - 4 \, \eta_\psi + 3 \left( d-m \right) \left( 1-z \right)
\right \rbrace  \left( \tilde V_S  + \frac{v_2 \, \tilde V_S^2 } {\gamma \, \varepsilon } \right) 
\nn
&
+ \, \big \lbrace  \lambda_1 \, \varepsilon - 4 \, \eta_\psi + 3 \left( d-m \right) \left( 1-z \right)
\big \rbrace \frac{ \left(   \frac{v_1 } {a } +  \frac{ v_5   \tilde V_S}{ \gamma }  + 
 \frac{ v_6 \tilde{V}_S } {a}\right) \tilde e } { N \, \varepsilon} 
 + \, \big \lbrace \left( \lambda_2 + \gamma \right) \varepsilon - 4 \, \eta_\psi + 3 \left( d-m \right) \left( 1-z \right)
\big \rbrace 
\frac{ \left(   \frac{v_3 } {a } +  v_4   \right) 
              \tilde e \, \tilde V_S } 
{ N \, \varepsilon}
= 0  \,, \nn
%
& \text{where }
\beta_V \equiv \frac{\partial \tilde{V}_S } {\partial \ln \mu}
= \beta_V ^{(0)} +  \beta_V ^{( 1 )}  \varepsilon  ,
\quad
 \tilde{\beta_e} \equiv \frac{\partial \tilde{e}  } {\partial \ln \mu} 
 = -\frac{m+1}{3} \,\varepsilon \, \tilde{e} + \mathcal{O } \left( \tilde{e}^2 \right).
\end{align}
Comparing the coefficients of the different powers of $\varepsilon $ on both sides, the solution takes the form:
\begin{align}
\frac{\partial \tilde{ V }_S }  { \partial l } =
\gamma \, \varepsilon  \tilde V_S 
- v_2   \tilde V_S^2
- \frac{(7-m^2) \, v_1 } { 6 \,N \, ( m+1) }\, \tilde e  
+ \, \Big[
\frac{  \left ( 5-m^2 -2m \right )\left  (3 \, v_5 - v_6 \right )   } 
{ m+1  }
+
(m-1)  \,  v_3   - 2\, (m+1)\, (  u_1 +  v_4  )
\Big]  \frac{\tilde e \,  \tilde{ V }_S } {6 \, N} \,,
\end{align}
upto $\mathcal{O} \left(  \tilde{e}^2, \tilde{ e} \, \varepsilon, \varepsilon^2  \right)$ in one-loop corrections.
This can be simplified as:
\bqa
\label{betav}
\frac{\partial \tilde{ V }_S }  { \partial l } =
\begin{cases}
\left(  \varepsilon-\frac{1 } { 2} \right)  \tilde V_S \, 
-v_2 \tilde V_S^2
-\frac{ v_1 } { 2 \, N} \, \tilde{e}
+\frac{    3 \, v_5-  4\,  v_4 - v_6  - 4 \, u_1  } { 6\, N} 
\, \tilde{e}\, \tilde V_S
& \mbox{ for } m=1\,,  \\
\gamma \, \varepsilon  \tilde V_S
-v_2 \tilde V_S^2
-\frac{  v_1 } { 6 \, N}   \, \tilde{e}
+\frac{  v_3 + 6\, v_4 - 3 \, v_5 + v_6 - 6 \, u_1
  } { 6 \, N} \, \tilde{e}\, \tilde V_S
& \mbox{ for } m=2 \mbox{ and } \gamma = \pm 1 \,.
\end{cases}
 \eqa
The RG flows for some representative values of $(d,m)$ have been plotted in Fig.~\ref{flows}.

\subsection{Fixed points of the beta functions}
\label{soln}

\begin{figure}
\centering
 \subfloat[][]{\includegraphics[width=0.45 \textwidth]{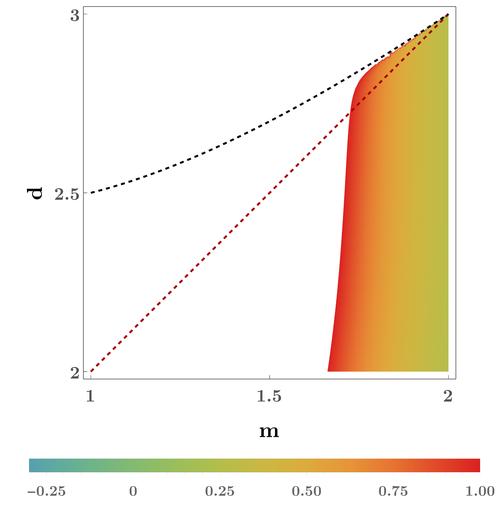}} 
\hspace{1 cm}
\subfloat[][]{\includegraphics[width=0.45 \textwidth]{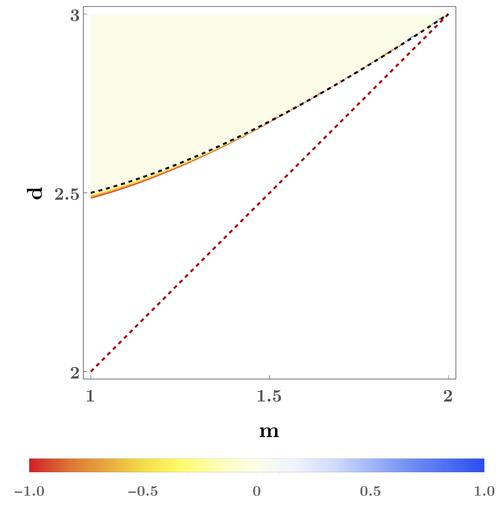}}
\caption{\label{roots}The contourplots of the two fixed points ($\tilde{ V }_S^*$) of $\beta_V$ as functions of $d$ and $m$. The intersection of the white areas in the two density-plots corresponds to regions where no perturbative fixed point exists. The dashed black(red) curve(line) represents $d =  d_c (d = \tilde d_c )$ in each plot.
}
\end{figure}

\begin{figure}
\centering
\includegraphics[width=0.45 \textwidth]{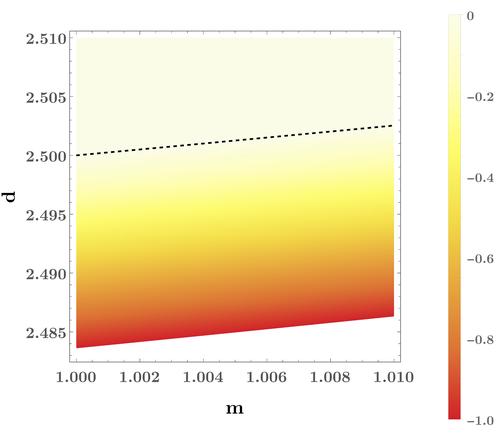}
\caption{\label{roots2}
The contourplot of the second root obtained from $\beta_V = 0$,  representing a fixed point value $\tilde{ V }_S^*$, as a function of $m$ and $d$. Here we have enlarged the region around $m = 1$. The dashed black line represents $d = d_c$.
}
\end{figure}

\begin{figure}
\centering
 \subfloat[][]{\includegraphics[width=0.44 \textwidth]{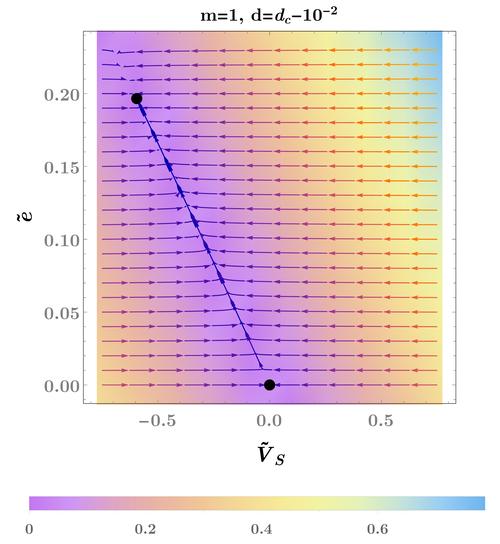} 
 { \label{flow1}}} \quad
\subfloat[][]{\includegraphics[width=0.44 \textwidth]{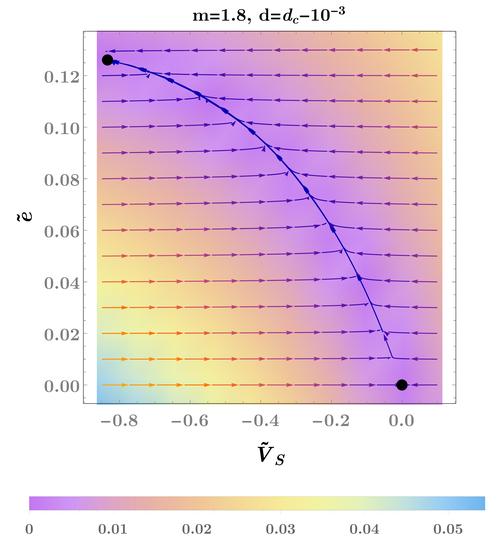}
{\label{flow2}}}\\
\subfloat[][]{\includegraphics[width=0.44 \textwidth]{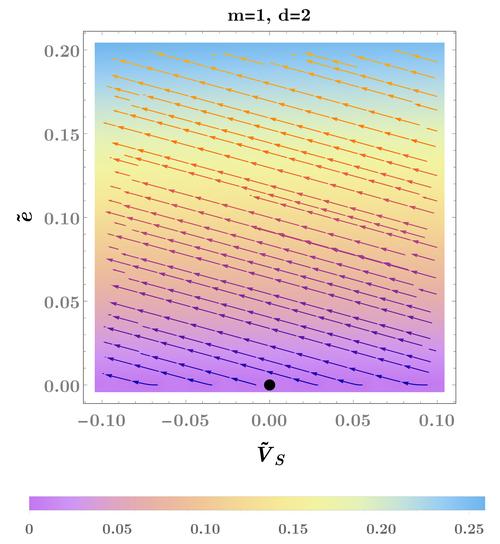} 
{ \label{flow3}}} \quad
\subfloat[][]{\includegraphics[width=0.44 \textwidth]{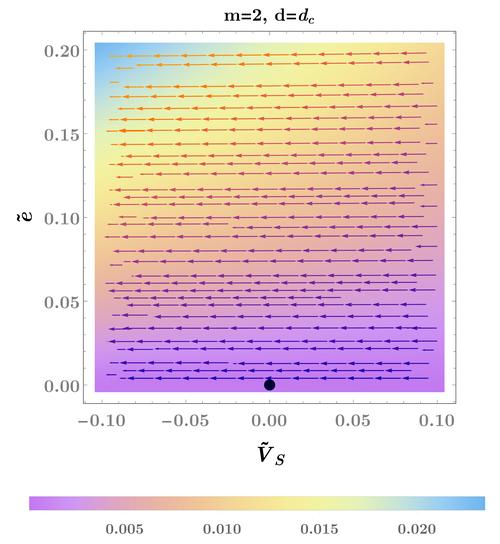}
{\label{flow4}}}
\caption{\label{flows}The representative RG flows in the various regions of the $(d,m)$-plane. The black dots represent the finite fixed points when they exist. The contour-shading conveys the magnitudes of the flow vector $\left(   \frac{\partial \tilde{ V }_S }  { \partial l } , \frac{\partial \tilde{ e}  }  { \partial l } \right)$ in different regions.}
\end{figure}


The fixed points for the beta functions can be found from the solutions of $\frac{\partial \tilde{ V }_S }  { \partial l } = 0 $ and $\frac{ \partial {\tilde e} }{ \partial l} = 0 $, using Eqs.~(\ref{betae}) and (\ref{betav}). The quadratic equation for the fixed point values of $  \tilde{ V }_S =  \tilde{ V }_S^* $ gives two roots. The contourplots of the values of these two roots as functions of $m$ and $d$ have been shown in Figs.~\ref{roots} and \ref{roots2}, restricting to the range $ |\tilde{ V }_S^*| < 1$. The reason for the restriction is that we must be careful not to go outside the window where the perturbation theory makes sense. The intersection of the white areas in the two subfigures of Fig.~\ref{roots} corresponds to regions where no perturbative solution exists and the system is eventually driven to superconducting state where $\tilde{ V }_S^* $ enters the strong coupling regime with large negative values (i.e., no longer within the perturbative regime), irrespective of the initial value of $ \tilde{ V }_S$ from which we start the RG flow. RG flows for some representative values of $m$ and $d$ have been plotted in Fig.~\ref{flows} in order to understand the stability of the fixed points. We observe the following:
\begin{enumerate}
\item At $(d=3, m=2)$, there is only finite solution given by $( \tilde{ V }_S^* =0, \tilde e^* = 0)$, corresponding to an IR unstable fixed point [cf. Fig.~\ref{flow4}]. This implies that for an initial $\tilde e  =0$, the system becomes superconducting only when the initial $\tilde V_S $ is negative, corresponding to the usual BCS instability in the Fermi liquid scenario. However, starting with any value of $\tilde e >0$ drives the system to the strong-coupling regime with large negative $\tilde V_S $, irrespective of the initial value of $ \tilde V_S$.  In other words, the system is unstable to superconductivity even for positive initial values of $\tilde{ V }_S $, in contrast with the case of Fermi liquids.\footnote{We would like to emphasize that the $(d=3, m=2)$ Fermi liquids are also unstable to Cooper pairing for an infinitesimally weak interaction regardless of its sign, but in some higher angular momentum channel ($\ell >0$) \cite{kohn}. For the Ising-nematic case under study, such an instability occurs for all angular momentum channels.} Hence, in this region, the non-Fermi liquid is unstable to superconducting instability with an enhancement brought about by the order parameter fluctuations.

\item
Let us first focus on the region near $(d=5/2, m=1)$, which has been magnified
in Fig.~\ref{roots2}. In a very thin slice below $d_c$ for $m = 1$, we find that there are two perturbative solutions, namely $( \tilde{ V }_S^* = -f_1, 
\, \tilde e^* = 
\frac{N \, \varepsilon } {u_1 } ) $ and $( \tilde{ V }_S^* = 0, 
\tilde e^* = 0 ) $, where $ 0<f_1<1 $ (e.g., Fig.~\ref{flow1}). In this region, the non-Fermi liquid is stable for a given initial $\tilde e > 0 $. However, if we switch our focus to the wide regions around region $(d=2, m=1)$, no perturbative solution for $\tilde{ V }_S^*$ exists and the system is unstable to superconductivity even for repulsive initial values of $\tilde{ V }_S $ (e.g., Fig.~\ref{flow3}). This is applicable for $m= 1$ and $d=d_c -\varepsilon$, with $\varepsilon  \gtrsim  0.017 $.
The Ising-nematic order parameter fluctuations thus lead to strong enhancement of superconducting instability in the physical region of interest, which is the region around $(d=2, m=1)$. But, unfortunately, our calculations are not perturbatively controlled in this region, as $\tilde e $ is not infinitesimally small for $\varepsilon \sim 1/2$, and we have derived our RG equations assuming that we are looking for poles in $1/ \varepsilon  $. Nevertheless, if we assume that we can continuously extrapolate to the physical case of $\varepsilon = 1/2$, we conclude that the pairing instability is strongly enhanced near the $(2+1)$-d Ising-nematic critical point and the destruction of quasiparticles is preempted by Cooper pairing, in agreement with Ref.~\cite{Max}. 
\vspace{2 cm}

\end{enumerate}

A few comments are in order regarding the enhanced superconductivity near $(d=3, m=2)$ and $(d=2, m=1)$. The Yukawa-like coupling $\tilde e$ gives rise to an effective attractive interaction for Cooper pairing and hence drives the system towards superconductivity even with an initially positive value of $ \tilde V_S$, as long as $\tilde e$ is non-zero, as the flow continues towards $\tilde V_S^*$ with large negative values. It can be shown that the energy scale at which this happens is always larger than the  non-Fermi liquid energy scale (\text{i.e.}, the scale where the non-Fermi liquid effects become appreciable). However, the magnitude of the pairing gap will depend on the initial values of the coupling constants. An analysis along these lines is very similar to that of Metlitski \textit{et al}. \cite{Max} and their estimate regarding the pairing gap holds after taking appropriate values of $\varepsilon$, namely $\varepsilon \simeq 0$ and  $\varepsilon = 1/2 $ for solving the $(d=3, m=2)$ and $(d=2, m=1)$ cases respectively.



\section{Summary and outlook}
\label{summary}

We have used RG to study potential superconducting instability induced by four-fermion interactions for Ising-nematic quantum critical points, in a perturbative expansion near the upper critical dimension $d_c (m)$ of the order parameter coupling with quasiparticles. Using the small parameter $\varepsilon = d_c(m) -d$, the RG flow equations and fixed points for both the effective order parameter coupling $\tilde e$ and BCS-channel coupling $\tilde V_S$ have been computed as functions of $d$ and $m$. We have found that for $(d=3, m=2)$, the flow towards the non-Fermi liquid fixed point is preempted by Cooper pairing, such that for an initial $\tilde e > 0 $, an attractive or repulsive four-fermion interaction in the BCS-channel eventually leads to superconductivity, enhanced by the coupling $\tilde e$. 
In a thin slice near $(d=5/2, m=1 )$, the non-Fermi liquid fixed point is a stable fixed point.
To reach the physical scenario with ($d=2, m=1$), we have to set $\varepsilon = 1/2$, which is not quite perturbatively controlled. However, if we extrapolate our results for $\varepsilon > 0.17 $ to $d=2$, then the $(2+1)$-d quantum critical point is expected to have strong enhancement of superconductivity by the order parameter fluctuations, such that even an initially repulsive $\tilde V_S$ will eventually flow to the strong-coupling regime with large negative values.

Let us compare our results with previous works. In Ref.~\cite{Max}, the authors considered one-dimensional Fermi surfaces. In the present work, our formalism could treat both $m=1$ and $m=2$ Fermi surfaces. The $(d=2, m=1)$ result agrees with that of Ref.~\cite{Max}. The $(d=3, m=2)$ result differs from the transverse gauge field case studied in Ref.~\cite{ips2}, where the non-Fermi liquid arising from the interaction with a $U(1)$ gauge field has been shown to give rise to pairing for angular momentum channels $\ell \geq 2$ only when the gauge coupling is higher than a threshold value. The $(d=3, m=2)$ Ising-nematic case shows such pairing instability for all angular momentum channels and for any non-zero coupling constant $\tilde e $. This scenario persists even for a non-zero four-fermion pairing potential $\tilde V_{S/A}$.

Lastly, although we have considered the simplest case of zero angular momentum and two ($N=2$) flavours of fermions, our analysis can be easily extended to a scenario involving a non-zero angular momentum channel with any value of $N$. In future work, we would like to develop similar formalism for fermions coupled to transverse gauge fields, and also for SDW and CDW quantum criticl points at which the critical bosons carry finite momenta.

\section*{Acknowledgments}
We thank Sung-Sik Lee for suggesting this problem and for numerous helpful discussions. We are also grateful to Krishnendu Sengupta for providing valuable inputs in improving the manuscript.
The research was supported by  NSERC and the Templeton Foundation. Research at the Perimeter Institute is supported in part by the Government of Canada  through Industry Canada, and by the Province of Ontario through the Ministry of Research and Information.


\appendix
\section{Computation of the Feynman diagrams proportional to $V_S^2$}
\label{v2}

The diagrams resulting in terms proportional to $V_S^2$ are shown in Fig.~\ref{fig:loop-4VV}.

For $\vec p_1= \vec p_3 $, the contribution from Fig.~\ref{fig:4vv1} is proportional to:
\begin{eqnarray}
&& I_1( \mathbf{P_1} - \mathbf{P_3})=
- \int dk \, \mbox{Tr} \big[ G_0 (k+ \mathbf{P_1} - \mathbf{P_3}) \, G_0 (k)\big] \nn
&&= 2 \int dk \, \frac { \mathbf{K} \cdot  \left ( \mathbf{K +P_3 -P_1} \right ) + \delta_k^2  }
{ \left ( \mathbf{K}^2   + \delta_k^2  \right )
\,  \big [   \left ( \mathbf{K +P_3 -P_1} \right )^2    + \delta_k^2   \big ] }
\, \exp\left ( - \frac{2 \, \mathbf{L}_{(k)}^2} { k_F } \right )  \,.
\end{eqnarray}
Shifting the dummy variable $k_{d-m} \rightarrow k_{d-m} + \mathbf{L}_{(k)}^2 $ and using Feynman parametrization, we obtain:
\begin{eqnarray}
&&I_1( \mathbf{P_1} - \mathbf{P_3})
= 2 \int dk \,
\int_0^1 dt \,  \frac { \mathbf{K} \cdot  \left ( \mathbf{K +P_3 -P_1} \right ) + k_{d-m}^2   }
{ \big [  (1-t )  \left ( \mathbf{K}^2   + k_{d-m}^2  \right )
+ t \left ( \mathbf{K +P_3 -P_1} \right )^2    + t \,k_{d-m}^2   \big ]^2 }  \, 
\exp\left ( - \frac{2 \, \mathbf{L}_{(k)}^2} { k_F } \right )
\nn
&&= 2  \int dk \,
\int_0^1 dt  \,  
\frac {  \mathbf{K}^2  - t \, (1-t )  \left ( \mathbf{P_3 -P_1} \right )^2 + k_{d-m}^2   }
{ 
\big [  \mathbf{K}^2   + k_{d-m}^2 
+ t  \, (1-t) \left ( \mathbf{  P_3 -P_1} \right )^2     \big ]^2
 }  \, 
\exp \left ( - \frac{2 \, \mathbf{L}_{(k)}^2} { k_F } \right ) .
\end{eqnarray}
In the last line, we have performed the variable shift $ \mathbf{K} \rightarrow \mathbf{K} - t  \left ( \mathbf{P_3 -P_1} \right ) $.
The remaining integrals can be computed in a straightforward manner to give:
\begin{eqnarray}
I_1( \mathbf{P_1} - \mathbf{P_3})
= \frac{  2^{1-2 d +\frac{m}{2} }  \, k_F^{m/2}  \, \pi^{1-\frac{d} {2} }   \, \sec \left ( \frac{ (d-m)\, \pi } {2} \right )
\, | \mathbf{  P_3 -P_1} |^{d-m-1}
} 
{ \Gamma \left ( \frac{ d-m} { 2 } \right )} \,,
\end{eqnarray}
which is logarithmically divergent at $d-m=1$.

However, for $\vec p_1 \neq \vec p_3 $, the contribution from Fig.~\ref{fig:4vv1} is proportional to:
\begin{eqnarray}
&& \tilde I_1( p_1 - p_3 )=
- \int dk \, \mbox{Tr} \big[ G_0 (k+ p_1  - p_3 ) \, G_0 (k)\big] 
=  - \beta_d \,  \frac{ |\vec P_1 - \vec  P_3 |^{d-m} \, k_F ^{ \frac{m-1}{2} }}{|\vec{L}_{(p_1 - p_3 )}|} 
\end{eqnarray}
where $\beta_d$ is defined in Eq.~(\ref{babos}). This integral is the same as appears in the computation of the one-loop self-energy $\Pi (q)$ for the bosonic propagator \cite{ips1}, and is convergent. So we must have $\vec p_1 = \vec p_3 $ to get a contribution to the RG equations.

For Fig.~\ref{fig:4vv2}, the contribution is proportional to:
\begin{eqnarray}
&& I_2( p_2 - p_3 ) =
- \int dk \, \mbox{Tr} \big[ G_0 (k+ p_2 - p_3 ) \, G_0 (k)\big] =
\begin{cases}
I_1( \vec P_2 - \vec  P_3 )
&\mbox{for} \quad 
  |\vec L_{(p_2  - p_3 )} | \ll \frac{|\vec P_2 - \vec  P_3 |} {\sqrt{k_F }} \,, \\
 \tilde I_1( p_2 - p_3 )
&\mbox{for} \quad 
  |\vec L_{(p_2  - p_3 )} | \gg \frac{|\vec P_2 - \vec  P_3 |} {\sqrt{k_F }}  \,.  \nn
\end{cases}
\end{eqnarray}
Performing the angular momentum decomposition, we get
\begin{eqnarray}
\label{api}
&& \int d \theta \, \theta^{m-1} I_2( p_2 - p_3 ) \nn
&=& \int_0^{ \frac{|\vec P_2 - \vec  P_3 |} {\sqrt{k_F }} }   d |\vec L_{(p_2  - p_3 )} | \,
\frac{   |\vec L_{(p_2  - p_3 ) }|^{m-1}   I_2( p_2 - p_3 )  }    {   k_F^{\frac{m}{2}}   }  
+
\int^{\infty}_{ \frac{|\vec P_2 - \vec  P_3 |} {\sqrt{k_F }} }   d |\vec L_{(p_2  - p_3 )} | \,
\frac{   |\vec L_{(p_2  - p_3 ) }|^{m-1}   I_2( p_2 - p_3 )  }    {   k_F^{\frac{m}{2}}   } \nn
&=& 
\Big \lbrace
\frac{  2^{1-2 d +\frac{m}{2} }    \, \pi^{1-\frac{d} {2} }   \, \sec \left ( \frac{ (d-m)\, \pi } {2} \right )
} 
{ m \, \Gamma \left ( \frac{ d-m} { 2 } \right )\, k_F^{\frac{m}{2}} } 
 +   \frac{ \beta_d  }
 {   m-1    } 
 \Big \rbrace 
\,  \frac{| \mathbf{  P_2 -P_3 } |^{d-1}}
 {  k_F^{\frac{m}{2}}  }\,,
\end{eqnarray}
which is suppressed by $1/k_F^m$ compared to the terms contributing to the beta functions.

In Fig.~\ref{fig:4vv3}, using the notation $q_2 \equiv  p_1+p_2 $, we find that near $d-m= 1$, the terms resulting from the loop integral can be simplified and are proportional to
\begin{eqnarray}
&&
\int \left ( \prod_{s=1}^4 {dp_s }  \right)
\,\delta^{(d+1)} (p_1+p_2-p_3-p_4 ) 
\times  \frac{ 1 - \delta_{j_1, j_2}   } 
{  \left ( \mathbf{K}^2   + \delta_{ k }^2  \right )
\,  \Big [   \left ( \mathbf{K - Q_2 } \right )^2    + \delta_{k -q_2 }^2  \Big ]  }  \nn
&&
\,\, \times \,  \sum_{j_1, j_2} 
\int d^{d-m} \vec K 
\, \Big [
\,
\big \lbrace \bar \Psi_{j_2} (p_3) \, \gamma_0 \, \Psi_{j_2} (p_2)\big \rbrace
\, \big \lbrace\bar \Psi_{j_1 } (p_4 ) \, \gamma_0 \, \Psi_{j_1 } (p_1 ) \big \rbrace
\, \vec K \cdot \left( \vec Q_2 - \vec K \right) \nn
&& \qquad \qquad \qquad \qquad  + \,
\big \lbrace \bar \Psi_{j_2} (p_3) \, \gamma_{d-m} \, \Psi_{j_2} (p_2)\big \rbrace
\, \big \lbrace\bar \Psi_{j_1 } (p_4 ) \, \gamma_{d-m} \, \Psi_{j_1 } (p_1 ) \big \rbrace
\, \delta_{k } \, \delta_{k - q_2} \,
\Big ] \nn
 && 
   + \mbox{ terms not contributing to Cooper pairing} \nn
&=& 2 
\int \left ( \prod_{s=1}^4 {dp_s }  \right)
\,\delta^{(d+1)} (p_1+p_2-p_3-p_4 ) 
\sum_{j_1, j_2}  \psi^\dagger _{+ ,j_2} (p_3 ) \, \psi^\dagger _{ - ,j_1 } ( - p_1 )
\, \psi_{-, j_1} (- p_4 ) \, \psi_{+ , j_2 } (- p_3  )
\,\left  ( 1- \delta_{j_1, j_2}  \right ) \nn
&& \qquad \times \,  I_3( q_2)
   + \mbox{ terms not contributing to Cooper pairing} \,,\nonumber
\end{eqnarray}
where
\begin{eqnarray}
I_3( q_2)=
\int d ^ {d-m}\vec K 
\, \frac{
\vec K \cdot \left( \vec Q_2 - \vec K \right) + \delta_{k} \,  \delta_{ k -q_2 }
}
{  \left ( \mathbf{K}^2   + \delta_{ k }^2  \right )
\,  \Big [   \left ( \mathbf{K - Q_2 } \right )^2    + \delta_{k - q_2}^2  \Big ]  } \exp \left ( - \frac{2 \, \mathbf{L}_{(k)}^2} { k_F } \right )
= I_2 (q_2 )\,,\nn
\label{api3}
\end{eqnarray}
which therefore do not contribute to the RG-flow.


\section{One-loop diagrams proportional to $ \tilde e  \,V_S$}
\label{v1l}

The first set of diagrams generating terms proportional to $e^2 \, V_S$ is shown in Fig.~\ref{fig:loop-4V}. We will use $M$ to denote the matrices belonging to the set $\lbrace \sigma_z ,     \mathbb{I}_{2\times 2} \rbrace$.

The term  arising from  Fig.~\ref{fig:4V1} is given by:
\begin{eqnarray}
  && t^1_{M} (p_1, p_2,p_3) \nn
&& =
-\frac{V_S \, (ie)^2  \, \mu^{x+d_v} }  { 4 \,  N} \left(1- \delta_{j_1, j_2}  \right)
\times \frac{2} {2 !} \nn
&& \quad \times 
 \int dk  \,
D_1(k)
\lbrace  \bar \Psi_{j_1 }(p_3)    M  \gamma_{d-m}  G_0(p_3-k)  G_0 (p_1-k)
  \gamma_{d-m} \Psi_{j_2 }(p_1) \rbrace 
\lbrace \bar \Psi_{j_2 }(p_4)   M  \Psi_{j_1 }(p_2)\rbrace \nn
&=&
-\frac{V_S \, e^2  \, \mu^{x+d_v} \left( 1-\delta_{j_1, j_2}  \right)} 
 {4 \, N} 
\lbrace  \bar \Psi_{j_1 }(p_3) M  \Psi_{j_2 }(p_1) \rbrace  \,
\lbrace \bar \Psi_{j_2 }(p_4)   M  \Psi_{j_1 }(p_2)\rbrace \, J_{1}   \,,
\end{eqnarray}
where
\begin{eqnarray}
J_{1}  &=& i^2
\int dk \, 
\gamma_{d-m}  \, G_0(p_3-k)  \,G_0(p_1-k) \, \gamma_{d-m}\,D_1(k)\nn
&=& \int dk \, 
\frac{
  \left( \vec{K} +\vec P_1 \right) \cdot   \left( \vec{K} +\vec P_3 \right)+   \delta_{k + p_1}^2   }
{ \lbrace \left( \vec{K} +\vec P_1 \right)^2+\delta_{k + p_1}^2 \rbrace  \,
\lbrace \left( \vec{K} +\vec P_3 \right)^2+\delta_{k + p_1 }^2 \rbrace
 \big[ {\vec{L}^2_{(k)}}  +   \tilde \alpha  \, \frac{  | \vec K|^{d-m}}{ | \vec{L}_{(k)}| } \big]}
  \,.\nn
\end{eqnarray}
Shifting $k_{d-m} \rightarrow ( k-p_1 )_{d-m} -\vec L_{(k+ p_1 ) }^2 $, $\vec K  \rightarrow \vec K - \vec P_3$, and performing the $ \vec L_{(k)}$-integral, we get
\bqa
\label{j1000}
J_{1}
&=& \frac{ | \csc \lbrace \frac{(m+1) \pi}{3} \rbrace |  \, \pi^{  \frac{m } {2} } }
 {3   \, \Gamma \left (   \frac{m} {2} \right) \, \tilde \alpha^{\frac{2-m } {3}}    }  
\int \frac{d^{d-m } \vec K \, d k_{d-m} } {(2\pi)^{d}} \, 
\frac 
{\vec K \cdot \vec {(K + Q_3) }    +  k_{d-m}^2   }
{ \lbrace \left(  \vec{K} +\vec Q \right)^2   +  k_{d-m}^2  \rbrace  \,
\lbrace  \vec{K}  ^2+  k_{d-m}^2 \rbrace \, | \vec{K - P_3 } |^{2 \beta    }}\nn
& \simeq &
 \frac{ | \csc \left ( \frac{ \delta \pi} {3} \right ) |     \, \pi^{ \frac{ m } {2} } 
   }
 {3 \,  \Gamma \left (    \frac{m} {2} \right) \, \tilde \alpha^{\frac{ \delta  } {3}}    } 
 \int_0^1  dt
\int \frac{d^ {d-m}  \vec K   \, d k_{d-m} }        {(2\pi)^{d}   \, | \vec{K  } |^{2 \beta     }} \, 
\frac 
{  \vec K \cdot \vec {(K + Q_3) }    +  k_{d-m}^2    }
{ 
 \lbrace  t \left(  \vec{K} +\vec Q_3 \right)^2   +   (1-t)  \,  \vec{K}  ^2 +  k_{d-m}^2 \rbrace^2
}
\Big [ 1 +  \frac{ \delta \, (d -2 + \delta )  \,\vec K \cdot \vec P_3  } {3 \, \vec K^2}  \Big ]\nn
& = &
\frac{ | \csc \left ( \frac{ \delta \pi} {3} \right ) |     \, \pi^{ 1+\frac{ m } {2} } 
   }
 {6 \,  \Gamma \left (    \frac{m} {2} \right) \, \tilde \alpha^{\frac{ \delta  } {3}}    } 
 \int_0^1  dt
\int \frac{d^ {d-m}  \vec K   }        {(2\pi)^{d}   \, | \vec{K  } |^{2 \beta     }} \, 
\frac 
{  \vec K \cdot \vec {(K + Q_3) }    +  t \left(  \vec{K} +\vec Q_3 \right)^2   +   (1-t)  \,  \vec{K}  ^2   }
{ 
 \lbrace  t \left(  \vec{K} +\vec Q_3 \right)^2   +   (1-t)  \,  \vec{K}  ^2   \rbrace^{\frac{3}{2}}
} \nn
&& \qquad \qquad \qquad \qquad \times \, 
\Big [ 1 +  \frac{ \delta \, (d -2 + \delta )  \,\vec K \cdot \vec P_3  } {3 \, \vec K^2}  \Big ] , 
\eqa
where
\beq
\vec Q_3= \vec P_1- \vec P_3,   \, \beta = \frac{(d-m) \, ( 2-m ) }  { 6 } ,\, m=2-\delta ,
\, d= m+ \frac{3}{ m+1} -\varepsilon \,.
\eeq
such that $\varepsilon = \epsilon \left(   1-  \frac{ \delta } {3}  \right)$.
We have expanded the expression in terms of small $\vec P_3$, which enables us to analyse the singularity structure
without any loss of generality.
Defining $\mathcal{D} = \vec K ^2 + t \, (1-y ) \,(1-t \, +  \, y \, t ) \, \vec Q_3 ^2 $, we get
\bqa
J_{1}
& = &
 \frac{ | \csc \left ( \frac{ \delta \pi} {3} \right ) |     \, \pi^{ 1+\frac{ m } {2}} \, \Gamma \left(\beta + \frac{3}{2}\right)
   }
 {6 \,  \Gamma \left (    \frac{m} {2} \right) \, \tilde \alpha^{\frac{ \delta  } {3}} 
 \,\Gamma(\beta ) \, \Gamma \left( \frac{3}{2}\right)    } 
 \int_0^1 dy
 \int_0^1  dt
\int \frac{d^ {d-m}  \vec K   }    {(2\pi)^{d}  } \, 
\frac 
{  2 \vec K ^2     + y \,  t \left(  1-2 \, t + 2 \, y \, t  \right)  \vec Q_3 ^2      }
{ 
 \mathcal{D} ^{\beta + \frac{3} {2} } 
} \,
y^{\beta -1 }   \sqrt{ 1-y  }   \nn
&& \qquad \qquad \qquad \qquad \qquad \qquad  \qquad  \times \, 
\Big [ 1 -  \frac{  \left( 2 \, \beta +  { 3}  \right)
\,t^2 \, (1-y )  \,  \vec Q_3 \cdot \vec P_3  } 
{   \mathcal{D}  }  \Big ]\nn
&=&
 \frac { | \csc \left ( \frac{ \delta \pi} {3} \right ) | 
 }
 {  6\,  \tilde{\alpha}^{ \frac{ \delta  } {3}  }
 \left( 2 \pi \right)^{ \frac{3}{3+\delta }-\delta }
 \, \Gamma  \left( 1 - \frac{\delta} {2}\right) \, \Gamma  \left(   \frac{3} {6-2 \, \delta }\right)
 }
 \frac{ 1}  {\varepsilon \left(   1-  \frac{ \delta } {3} \right) |\vec Q_3 |^{\varepsilon \left(   1-  \frac{ \delta } {3} \right ) }} + \mathcal{O} (1)\,.
\nn\eqa
This is logarithmically divergent at $d = d_c $.
Near $(d =3, m=2)$, we expand in small delta such that 
\bqa
J_{1}
& = &
\frac{1} {4 \, \pi^2 }  \frac{ 1}  {\delta \, \varepsilon \left(   1-  \frac{ \delta } {3} \right )
 |\vec Q_3 |^{\varepsilon \left(   1-  \frac{ \delta } {3} \right ) }} + \mathcal{O} (1)\,.
\eqa
This indicates there is a $\log^2$ divergence. We note that there is no $\vec P_3 $-dependent term divergent in $\frac{1} {\delta }$ or $\frac{1} 
{ \varepsilon \left(   1-  \frac{ \delta } {3} \right )} $. In fact, the leading order term dependent on $\vec P_3 $ goes as $ 
\frac{ 1 - 2 \ln 2 } {    4 \, \pi^{ 5 } }
 \frac{ \vec Q_3 \cdot \vec P_3 } 
  { |\vec Q_3 |^{2 + \varepsilon  \left(   1-  \frac{ \delta } {3} \right )   }} $.

The term arising from Fig.~\ref{fig:4V2} has a divergent structure identical to Fig.~\ref{fig:4V1}. Depending on $m$, these two diagrams are of either higher order than or same order as the $V_S^2 $ terms. Near $(d=5/2, m=1 )$, these have only a pole in $\varepsilon$ (and not in $\delta$) and hence are of leading order. On the other hand, near $(d=3, m=2 )$, these have two poles (one in $\varepsilon$ and one in $\delta$) and hence are one order higher than the $V_S^2$ terms.
In this second scenario, namely near $(d=3, m=2 )$, these two diagrams must then be considered in conjunction with those consisting of the counterterm vertices generated from the divergence of the angular momentum decomposition of Fig.~\ref{bcstree}, as shown in Fig.~\ref{evc}. In Sec.~\ref{logsq}, we have shown how these terms cancel out around this region.

The term arising from  Fig.~\ref{fig:4V3} is given by:
\begin{eqnarray}
&& t^3_{M} (p_1, p_2,p_3) \nn
&& =
-\frac{V_S \, (ie)^2  \, \mu^{x+d_v} }  { 4 \,  N} 
\left( 1-\delta_{j_1, j_2}  \right)  \nn
&& \quad \times
 \int dk  \,
D_1(k-p_3)
\lbrace  \bar \Psi_{j_1 }(p_3)    \gamma_{d-m}  G_0(k) M \Psi_{j_2 }(p_1) \rbrace 
\lbrace \bar \Psi_{j_2 }(p_4)  \gamma_{d-m}  G_0(p_1 + p_2 - k) M  \Psi_{j_1 }(p_2)\rbrace \nn
& & \simeq
-\frac{V_S \, e^2  \, \mu^{x+d_v} }  {4 \, N} \left( 1-\delta_{j_1, j_2}  \right) 
\lbrace  \bar \Psi_{j_1 }(p_3) M  \Psi_{j_2 }(p_1) \rbrace  \,
\lbrace \bar \Psi_{j_2 }(p_4)   M  \Psi_{j_1 }(p_2)\rbrace \, J_{3}   \,,
\end{eqnarray}
where we have dropped terms irrelevant for superconductivity such that
\begin{equation}
J_{ 3 } 
= -\int dk \, 
\frac{
  \vec{K}  \cdot   \left( \vec{K} -\vec P_1 -\vec P_2  \right) -  \delta_k \, \delta_{k - p_1-p_2}  }
{ \lbrace  \vec{K} ^2+\delta_{k  }^2 \rbrace  \,
\lbrace \left( \vec{K} -\vec P_1-\vec P_2 \right)^2+\delta_{ k - p_1-p_2  }^2 \rbrace
 \big[ {\vec{L}^2_{(k-p_3)}}  +   \tilde \alpha  \, \frac{  | \vec K-\vec P_3|^{d-m}}{ | \vec{L}_{(k-p_3)}| } \big]} 
 = 0 \,,
\end{equation}
on performing the $k_{d-m}$ integral. Similarly, the contribution from Fig.~\ref{fig:4V4} also vanishes.

Next we consider Fig.~\ref{fig:4V5}, again only keeping terms relevant for superconductivity. This contributes as:
\begin{eqnarray}
  && t^5_{M} (p_1, p_2,p_3) \nn
&& =
-\frac{V_S \, (ie)^2  \, \mu^{x+d_v} }  { 4 \,  N} \left( 1-\delta_{j_1, j_2}  \right) \nn
&& \quad \times
 \int dk  \,
D_1(k )
\lbrace  \bar \Psi_{j_1 }(p_3) M   G_0(p_1 - k)   \gamma_{d-m}  \Psi_{j_2 }(p_1) \rbrace 
\lbrace \bar \Psi_{j_2 }(p_4)  \gamma_{d-m}  G_0(p_4 - k) M  \Psi_{j_1 }(p_2)\rbrace \nn
& & \simeq
-\frac{V_S \, e^2  \, \mu^{x+d_v} }  {4 \, N} \left( 1-\delta_{j_1, j_2}  \right)
\lbrace  \bar \Psi_{j_1 }(p_3) M  \Psi_{j_2 }(p_1) \rbrace  \,
\lbrace \bar \Psi_{j_2 }(p_4)   M  \Psi_{j_1 }(p_2)\rbrace \, J_{5 }   \,,
\end{eqnarray}
where
\begin{eqnarray}
J_{5} 
&=&  \int dk \, 
\frac{
 \left(  \vec{K-P_1 } \right) \cdot   \left( \vec{K} -\vec P_4   \right) 
 +  \delta_{k-p_1} \, \delta_{k - p_4 }  }
{ \lbrace   \left(  \vec{K-P_1 } \right)   +\delta_{k -p_1 }^2 \rbrace  \,
\lbrace \left( \vec{K} -\vec P_4 \right)^2+\delta_{ k - p_4  }^2 \rbrace
 \big[ {\vec{L}^2_{(k )}}  +   \tilde \alpha  \, \frac{  | \vec K |^{d-m}}{ | \vec{L}_{(k )}| } \big]} \nn
&= & \int dk \, 
\frac{
  \vec{K}  \cdot   \left( \vec{K} + \vec P_1 - \vec P_4  \right) +  \delta_{k+p_1} \, \delta_{k + p_2 }  }
{ \lbrace  \vec{K} ^2+\delta_{k +p_1 }^2 \rbrace  \,
\lbrace \left( \vec{K} +\vec P_1 -\vec P_4 \right)^2+\delta_{ k + p_2  }^2 \rbrace
 \big[ {\vec{L}^2_{(k  )}}  +   \tilde \alpha  \, \frac{  | \vec K + \vec P_1 |^{d-m}}
 { | \vec{L}_{(k  )}| } \big]} \,.\nonumber
 \end{eqnarray}
Setting $ \vec p_1 (= \vec p_3) = 0 $ and $\vec P_1 = \vec P_4 $ for simplification, which is valid for extracting the divergent term, we get:
\begin{eqnarray}
J_{ 5 }
 &= & \int d k \, 
\frac{
  \vec{K}^2   +  \delta_k \, \delta_{k +  p_2}  }
{ \lbrace  \vec{K} ^2+\delta_{k  }^2 \rbrace  \,
\lbrace   \vec{K}^2+\delta_{ k  + p_2  }^2 \rbrace
 \big[ {\vec{L}^2_{(k )}}  +   \tilde \alpha  \, \frac{  | \vec K + \vec P_1 |^{d-m}}{ | \vec{L}_{(k )}| } \big]} \nn
 &=&  \frac{ 1 } {\pi  \, (2 \pi)^d}
 \int 
\frac{  d^{d-m} \vec K \, d^{m} \vec L_{(k)}   \, |\vec K|  }
{ \lbrace   \vec{K} ^2 +  \vec L_{(k)}^2 \,  \vec L_{(p_2)}^2 \, \cos^2 \theta_k \rbrace  \,
 \big[ {\vec{L}^2_{(k )}}  +   \tilde \alpha  \, \frac{  | \vec K + \vec P_1 |^{d-m}}{ | \vec{L}_{(k )}| } \big]} \,. \nonumber
\end{eqnarray}
In the last step, we have ignored the terms involving $ \delta_{p_2  } $ in the denominator, and used $\theta_k $ to denote the angle between $\vec L_{(k)}$ and $\vec L_{( p_2 )}$.

In order to proceed further and obtain an analytic expression, we need to examine which terms in the integral lead to singularity.
If $k = ( {\bf K}, k_{d-m}, \vec L_{(k)} )$ denotes the momentum flowing through a boson propagator within a loop diagram,
such that $|{\bf K}|$ is order of $\Lambda$,
the typical momentum carried by a boson along the tangential direction of the Fermi surface is given by
$ |{\vec{L}}_{(k)} |^3 \sim 
\tilde \alpha  \, \Lambda^{d-m} $. 
For a fixed value of $\tilde{e} \sim \mathcal{O} (\varepsilon)$, we can neglect the first term for $\tilde \alpha^{\frac{1}{3}}  \, \Lambda^{ \frac{d-m}{3} } \, |\vec L_{(p_2)} | \gg \Lambda $, and the second term for $ \alpha^{\frac{1}{3}}  \, \Lambda^{ \frac{d-m}{3} } \, |\vec L_{(p_2)} | \ll \Lambda $, in the
$\lbrace   \vec{K} ^2 +  \vec L_{(k)}^2 \,  \vec L_{(p_2)}^2 \, \cos^2 \theta_k \rbrace$ factor of the denominator \cite{ipshigh}. This gives us
\begin{eqnarray}
 J_{ 5 }
 &= &
 \begin{cases}
   \int d^{d-m}\vec K \,
 \frac{  \pi^{\frac{7-2d -\delta } {2}}   \, \csc \left( \frac{\delta-1}{3} \pi \right)
 \sec \left( \frac{\delta \pi }{2}   \right) \, |\vec K | } 
 { 3\times 2^{ d-1+\delta} \,  \Gamma(-\delta )\, \Gamma \left( \frac{ 1 + \delta } {2} \right)
\, \vec L_{(p_2)}^2  \, |\vec K + \vec P_1 |^{  \frac{ (2 + \delta)\, (d-2 + \delta ) } { 3 }} 
 \, \tilde \alpha^{\frac{2+ \delta} {3}} 
  } \qquad &\mbox{for} \quad 
   \tilde \alpha^{\frac{1}{3}}  \, \Lambda^{ \frac{d-m}{3} } \, |\vec L_{(p_2)} | \gg \Lambda \,,  \\
 \int d^{d-m}\vec K \,
 \frac{  \pi^{\frac{4-2d +m } {2}}   \, \csc \left( \frac{m+1}{3} \pi \right)
  } 
 { 3\times 2^{ d-1 }  \, \Gamma \left( \frac{ m } {2} \right)
 \, |\vec K|  \, |\vec K + \vec P_1 |^{  \frac{ (d-m)\, ( 2-m ) } { 3 }} 
 \, \tilde \alpha^{\frac{2 - m } {3}} 
  } \qquad &\mbox{for} \quad  \tilde \alpha^{\frac{1}{3}}  \, \Lambda^{ \frac{d-m}{3} } \, |\vec L_{(p_2)} | \ll \Lambda \,, 
 \end{cases}\nonumber
 \end{eqnarray}
where $\delta = 2-m$.
Finally, performing the $\vec K$ integrals, the leading order terms are found to be:
\begin{eqnarray}
J_{ 5 } =
\begin{cases}
    \frac {\sqrt{3} \,  |\vec P_1 |^{\frac{ 4-
\left(  1- \frac{\delta } {3}  \right)  \, \varepsilon -2 \delta+ 
\left(  1- \frac{\delta } {3}  \right)  \, \varepsilon   \, \delta } { 3 -\delta }}   } 
 { 4   \, \tilde \alpha^{\frac{2+ \delta} {3}} \, \vec L_{(p_2)}^2
  } \, \delta   
  \qquad &\mbox{for} \quad    \tilde \alpha^{\frac{1}{3}}  \, \Lambda^{ \frac{d-m}{3} } \, |\vec L_{(p_2)} | \gg \Lambda  \,,  \\
  \frac{
 \pi^{2- \frac{m}{2}  -  \frac{3 } {2\, (m+1) }}   \, \csc \left( \frac{m+1}{3} \pi \right)}
 {  2^{ \frac{m\,(2-m)} {m+1} }\, (m+1) \, \Gamma \left( \frac{ m } {2} \right)
 \, \Gamma \left( \frac{ 3 } {2 \, (m+1) } \right) \, \tilde \alpha^{\frac{2 - m } {3}} 
  }  
  \frac{1}{2 \, |\vec P_1  |^{\frac{(m+1) \, \varepsilon} {3 }} \, \varepsilon } 
  \qquad &\mbox{for} \quad   \tilde \alpha^{\frac{1}{3}}  \, \Lambda^{ \frac{d-m}{3} } \, |\vec L_{(p_2)} | \ll \Lambda  \ \,.
\end{cases}
 \end{eqnarray}
Now we need to carry out the angular momentum decomposition to get
\begin{eqnarray}
&& e^2 \int  d\theta \, \theta^{ m-1 } J_{ 5 } =
e^2\int_0^{\frac{\Lambda^{\frac{3-d+m}{3}}  }
{ \tilde  \alpha^{\frac{1 } {3} }  }} 
d |\vec{L}_{ (p_2)}| \,\frac{ |\vec{L}_{ (p_2)}|^{ m-1 } J_{ 5 } }
{ k_F^{\frac{m}{2}}  }
+ 
e^2 \int^{\infty}_{\frac{\Lambda^{\frac{3-d+m}{3}}  }
{ \tilde  \alpha^{\frac{1 } {3} }  }} 
d |\vec{L}_{ (p_2)}| \,\frac{ |\vec{L}_{ (p_2)}|^{ m-1 } J_{ 5 } }
{ k_F^{\frac{m}{2}}  }\nn
&& =
 \frac{
 \pi^{2- \frac{m}{2}  -  \frac{3 } {2\, (m+1) }}   \, \csc \left( \frac{m+1}{3} \pi \right)}
 {  2^{ \frac{2+ m^2 } {m+1} }\, (m+1) \, \Gamma \left( \frac{ m +2 } {2} \right)
 \, \Gamma \left( \frac{ 3 } {2 \, (m+1) } \right) \, \tilde \alpha^{\frac{2 } {3}} 
  }  
  \frac{ e^2 \, \Lambda^{\frac{ (3-d+m) \, m } {3} } }
  {  |\vec P_1  |^{\frac{(m+1) \, \varepsilon} {3 }} \,   k_F^{\frac{m}{2}} \, \varepsilon} 
  +
 \frac {\sqrt{3} \,  |\vec P_1 |^{\frac{ 4-
\left(  1- \frac{\delta } {3}  \right)  \, \varepsilon -2 \delta+ 
\left(  1- \frac{\delta } {3}  \right)  \, \varepsilon   \, \delta } { 3 -\delta }}   } 
 { 4   \, \tilde \alpha^{\frac{2 } {3}} 
  } 
 \frac{ e^2 }  { \Lambda^{\frac{ (3-d+m) \, \delta } {3} } \, k_F^{\frac{m}{2}}   } 
  ,\nn 
\end{eqnarray}
which is suppressed by an overall factor of $\tilde e^{\frac{1}{m+1}} / k_F^{\frac{m^2} {m+1}} $.
Hence, this diagram is not relevant for the RG-equations.
It is easy to see that Fig.~\ref{fig:4V6} has the same singularity behaviour as Fig.~\ref{fig:4V5}.

\subsection{Cancellation of the terms proportional to $\tilde e \, \tilde V_S $ near $m=2$ }
\label{logsq}

The $\log^2$ divergences in the diagrams in Fig.~\ref{fig:loop-4V} near $m=2$ need to be examined carefully, as potentially they can give rise to non-local terms in the renormalized action.
For $m=2 -\delta$ and $d = d_c - \varepsilon $, the diagram in  Fig.~\ref{fig:4V1}, for example, diverges as
\begin{eqnarray}
-\frac{2 \, \tilde e \, \tilde  V_S} {4  \, N } 
J_1
=
\frac{  \tilde e \, \tilde  V_S} { 8 \, \pi^2 \, N } 
\Big [ -\frac{ 1 }
{  \delta \, \varepsilon  \left(   1-  \frac{ \delta } {3} \right )}
+\frac{ \ln \left( \frac{ |\vec P_1 - \vec P_3 | } {\mu}\right ) }  {\delta}
+\frac{ \gamma_{E} }  {  \varepsilon  \left(   1-  \frac{ \delta } {3} \right )  }
+\frac{ \gamma_{E} }  {\delta }
\Big ] ,
\end{eqnarray}
where $\gamma_E = 0.577 $ is the Euler-Masheroni constant.
The above must be considered in conjunction with the diagram in Fig.~\ref{eV1}, which gives a divergent term equal to
\bqa
\frac{2 \times 2} {4 }
\frac{ \tilde e \, \tilde V_S}  {N  }
\frac{1}
{ 2^{ \frac{ 3m} {2} } \, \pi^{1+\frac{m} {2}}   }
\Big [ \frac{ 1 -   \gamma \, \varepsilon   \, 
\ln\left( \frac{ |\vec P_1 - \vec P_3 | } {\mu}\right ) }   
{\delta \, \gamma \, \varepsilon} 
- \frac{ \gamma_{E} }  {\gamma \, \varepsilon }
+ \frac{ \gamma_{E} }  {\delta }  \Big]
=
\frac{ \tilde e \, \tilde  V_S} {8 \, \pi^2   N  } 
\Big [\frac{ 1}  { \delta \, \gamma \, \varepsilon} 
-   \frac{ \ln \left( \frac{ |\vec P_1 - \vec P_3 | } {\mu}\right ) } {\delta}
-
\frac{ \gamma_{E} }  {\gamma \, \varepsilon }
+\frac{ \gamma_{E} }  {\delta } \Big] .\nn
\eqa
Clearly, the coefficients of $ \ln \left ( \frac{ |\vec P_1 -\vec P_3| }  { \mu } \right )$ cancel and there is no dangerous non-local term in the renormalized action. Moreover, the $log^2$ divergence at $(d=3, m=2)$ also cancels out (with $\gamma=1$), as expected in the minimal subtraction scheme.

\section{Treatment of the terms proportional to $\tilde e $ in the Wilsonian RG language}
\label{wil}

From our boson propagator $D_1(q)$, it is clear that the instantaneous interaction is mediated by bosons with momenta $|\mathbf{L}_{(q)}|\gtrsim \lambda_t \equiv e^{2/3} \beta_d^{1/3} \Lambda^{\frac{d-m+x} {3}} k_F^{\frac{m-1} {6}} $. In the words of Son \cite{son}, \textit{part of the non-instantaneous interaction becomes instantaneous} when RG is carried out with $ e^{-\frac{(4-m) \, l}{6} }\, \lambda_t < |\mathbf{L}_{(q)}  | < \lambda_t $. Here we note that since we are using the $dressed$ bosonic propagator, this is the correct interval to be considered. On the other hand, in the analysis of Metlitski \textit{et al}. \cite{Max}, since the cut-off in the tangential direction is proportional to $\sqrt{ \Lambda \, k_F}$, RG is carried out with $ e^{\frac {-l} {2} } \, \sqrt{ \Lambda \, k_F} < |\mathbf{L}_{(q)}  | < \sqrt{ \Lambda \, k_F} $. These two approaches match for $m=1$. It is more logical to integrate out the instantaneous part of the boson propagator with $|\mathbf{L}_{(q)}|$ in the appropriate range obtained following Son's approach \cite{son}. Moreover, for $m > 1$, there is an already non-trivial $k_F$ dependence from UV-IR mixing arising due to inter-patch coupling, caused by the interactions with the massless boson \cite{ips1}. We conclude that in the presence of the four-fermion interactions, the UV-IR mixing becomes manifest even for $ m= 1 $.

Setting $\mathbf{P_1}=\mathbf{P_2}$ gives  the ``instantaneous'' contribution in the BCS-channel with angular momentum $\ell$, which can be expressed as:
\begin{eqnarray}
t_{ee} (l, \ell)
&=&
\begin{cases}
\frac{e^2 \, \Lambda^x} {2 \, N} \times 2  \int_{\theta >0} 
   \frac{ d \theta} { 2 \, \pi} \, \frac{  e^{-i \ell \, \theta} } { \mathbf{L}_{(q)} ^ 2 } 
& \mbox{ for } m=1\,, \\
 \frac{e^2 \, \Lambda^x} {4 \, N} \int d \theta \, \frac{  \sin \theta \, P_{\ell}  \left ( \cos \theta \right ) } { \mathbf{L}_{(q)} ^ 2 } 
& \mbox{ for } m=2 \,.
\end{cases}
\end{eqnarray}
Using $ \epsilon= m+1 - d $, we get
 \begin{eqnarray}
t_{ee} (l,\ell)
&\simeq&
\begin{cases}
  \frac{e^2 \, \Lambda^x}  { 2\, N \, \pi}
\int_{ e^{-\frac{  l}{2} }\, \lambda_t }^{ \lambda_t  } d  |\mathbf{L}_{(q)}|
\, \, \frac{   1 } {   \sqrt {k_F} \,\mathbf{L}_{(q)}^2 }
=   \frac{e^2 \, \Lambda^x} { 2 \,N \,  \pi \,  \sqrt{ k_F} \,\lambda_t} \frac{ l}{ 2 } 
=  \frac{\tilde e \, \Lambda^\epsilon } {  N \,\sqrt{ k_F} \,\beta_d^{1/3}} \frac{ l}{ 4 \, \pi } 
& \mbox{ for } m=1 \,,\\
 \frac{e^2 \, \Lambda^x} {  4 \, N}
\int_{ e^{-\frac{  l}{3} }\, \lambda_t }^{ \lambda_t  } d  |\mathbf{L}_{(q)}|
\, \, \frac{   P_{\ell}   ( 1 ) } { k_F  \,| \mathbf{L}_{(q)}  | }
=    \frac{\tilde e  \, \Lambda^\epsilon} {4 \,N \,  k_F } \frac{ l}{ 3 } 
& \mbox{ for } m=2 \,,\nonumber
 \end{cases}
 \\
\end{eqnarray}
which is independent of the angular momentum channel $\ell$.

\bibliography{sc}

\end{document}